\documentclass[twocolumn,showpacs,showkeys,preprintnumbers,superscriptaddress,amsmath,floatfix,amssymb,secnumarabic]{revtex4}

\usepackage[colorlinks=true]{hyperref}
\usepackage{graphicx}

\newcommand{\comment}[1]{}

\newcommand{\lr}[1]{ \left( #1 \right) }
\newcommand{\lrs}[1]{ \left[ #1 \right] }

\newcommand{\vev}[1]{ \langle \, #1 \, \rangle }

\newcommand{\tr}{ {\rm tr} \, }
\newcommand{\re}{ {\rm Re} \, }

\renewcommand{\Im}{ {\rm Im} \, }

\newcommand{\ket}[1]{ \, | #1 \rangle }
\newcommand{\bra}[1]{ \langle #1 | \, }

\newcommand{\const}{ {\rm const}}
\renewcommand{\det}[1]{ {\rm det} \left( #1 \right) }
\newcommand{\absval}[1]{ \left| #1 \right| }

\newcommand{\sign}{ {\rm sign} \,  }

\newcommand{\dslash}[1]{ #1 \!\!\!/}

\begin{document}
\sloppy

\title{Anomalous transport with overlap fermions}

\author{P. V. Buividovich}
\email{pavel.buividovich@physik.uni-regensburg.de}
\affiliation{Institute of Theoretical Physics, University of Regensburg, D-93053 Germany, Regensburg, Universit\"{a}tsstra{\ss}e 31}

\date{February 28th, 2014}
\begin{abstract}
 Anomalous correlators of vector and axial currents which enter the Kubo formulae for the chiral magnetic and the chiral separation conductivities are explicitly calculated for free overlap fermions on the lattice. The results are confronted with continuum calculations in the finite-temperature regularization, and a subtle point of such regularization for chiral magnetic conductivity related to the correct counting of the chiral states is highlighted. In agreement with some previous claims in the literature, we find that in a lattice regularization which respects gauge invariance, the chiral magnetic conductivity vanishes. We point out that the relation of anomalous transport coefficients to axial anomaly is nontrivial due to the non-commutativity of their infrared limit and the Taylor expansion in baryon or chiral chemical potential. In particular, we argue that the vector and axial Ward identities fix the asymptotic behavior of anomalous current-current correlators in the limit of large momenta. Basing on the work of Knecht \emph{et al.} on the perturbative non-renormalization of the transverse part of the correlator of two vector and one axial currents, we demonstrate that the relation of the anomalous vector-vector correlator to axial anomaly holds perturbatively in massless QCD but might be subject to non-perturbative corrections. Finally, we identify kinematical regimes in which the anomalous transport coefficients can be extracted from lattice measurements.
\end{abstract}

\maketitle

\section{Introduction}
\label{sec:introduction}

 Anomaly-related transport properties of systems of chiral fermions have become recently a subject of intense theoretical and experimental studies. Anomalous transport phenomena of particular interest for the physics of heavy-ion collisions are the Chiral Separation Effect (CSE) \cite{Metlitski:05:1, Son:06:2} and the Chiral Magnetic Effect (CME) \cite{Kharzeev:08:2}. The Chiral Separation Effect is the generation of an axial current $j^A_i$ along the magnetic field $B_i$ at nonzero chemical potential $\mu_V$:
\begin{eqnarray}
\label{CSEClassicExpression}
 j^A_i = \sigma_{CSE} \, B_i, \quad \sigma_{CSE} = \frac{1}{2 \pi^2} \mu_V  .
\end{eqnarray}
The Chiral Magnetic Effect is the generation of an electric current $j^V_i$ along the magnetic field at nonzero chiral chemical potential $\mu_A$:
\begin{eqnarray}
\label{CMEClassicExpression}
 j^V_i = \sigma_{CME} \, B_i, \quad \sigma_{CME} = \frac{1}{2 \pi^2} \mu_A .
\end{eqnarray}
For simplicity in this paper we assume that the numbers $N_c$ and $N_f$ of fermion colours and flavours and also the fermion charge are all equal to one. These numbers enter all the expressions which we use in this work as simple pre-factors and can be restored, if necessary, in a completely straightforward way.

 While the chemical potential $\mu_V$ couples to the total charge $q_R + q_L$ of both right- and left-handed fermions, the chiral chemical potential $\mu_A$ couples to the difference $q_R - q_L$. Both Chiral Magnetic and Chiral Separation effects combine into a novel type of excitation of a plasma of chiral fermions in an external magnetic field - the Chiral Magnetic Wave \cite{Kharzeev:10:1, Kharzeev:11:1}. All these effects should lead to specific anisotropies and correlations in the distributions of charged particles produced in heavy ion collisions \cite{Kharzeev:06:1, Kharzeev:08:1, Kharzeev:11:2}. In a recent work \cite{Hongo:13:1} first numerical simulations of hydrodynamical evolution in heavy-ion collisions which took into account anomalous transport were performed and some specific experimental signatures of the importance of anomalous transport phenomena were pointed out. Even though the experimental evidence for these effects from heavy-ion collision experiments is somewhat controversial at present (see e.g. \cite{Skokov:13:1}), anomalous transport phenomena might also be interesting in other areas of physics, for example, for the recently discovered Weyl semimetals \cite{Wan:11:1, Basar:13:1, Landsteiner:13:1, Chernodub:13:1}.

 In contrast to the conventional transport phenomena such as ohmic conductivity or viscous flow, anomalous transport is a property of equilibrium state of quantum systems and does not lead to any dissipation of heat. For this reason they are also ideally suited for lattice studies. Indeed, since within the linear response approximation anomalous transport coefficients can be extracted from static correlators of axial and vector currents and the energy-momentum tensor \cite{Landsteiner:11:2, Landsteiner:12:1}, tedious analytic continuation of lattice data from Euclidean to real (Minkowski) time which is typically a source of large uncertainties and systematical errors turns out to be unnecessary.

 In the original works \cite{Metlitski:05:1, Son:06:2} and \cite{Kharzeev:08:2} (see also \cite{Nielsen:83:1, Vilenkin:80:1} for some earlier derivations, and \cite{Hong:10:1} for a later work on non-renormalizability of CME) it was argued that the values (\ref{CSEClassicExpression}) and (\ref{CMEClassicExpression}) of the transport coefficients $\sigma_{CSE}$ and $\sigma_{CME}$ (which we will call the chiral separation conductivity and the chiral magnetic conductivity, respectively) are related to the axial anomaly and thus do not receive any corrections due to interactions. Somewhat later it has also been realized that the values of $\sigma_{CSE}$ and $\sigma_{CME}$ can be strongly constrained based on purely thermodynamical arguments if one requires that the divergence of an entropy current in ``anomalous hydrodynamics'' is non-negative \cite{Son:09:1, Sadofyev:10:1, Sadofyev:10:2, Zakharov:12:1, Jensen:12:1, Jensen:12:2, Banerjee:12:1}. Anomalous transport coefficients (\ref{CSEClassicExpression}) and (\ref{CMEClassicExpression}) are also reproduced in classical kinetic theory, where the quantum anomaly can be incorporated in terms of the nontrivial flux of the momentum-space Berry curvature through the Fermi surface \cite{Son:12:2, Stephanov:12:1}.

 However, some further analytic and numerical studies \cite{Fukushima:10:1, Yamamoto:11:1, Yamamoto:11:2} indicated that both the Chiral Magnetic Effect and the Chiral Separation Effect might still receive corrections in interacting field theories. It was recently stressed in \cite{Jensen:13:1} that if one couples dynamical gauge fields to the currents entering the anomalous correlators, the anomaly equations are no longer restrictive enough to completely fix the anomalous transport coefficients. A direct calculation \cite{Miransky:13:1} of radiative corrections to the chiral separation conductivity confirms this statement.

 Chiral magnetic and chiral separation conductivities has also been actively studied in holographic models \cite{Kalaydzhyan:11:1, Yee:09:1, Bergman:09:1, Rubakov:10:1, Gynther:10:1, Gorsky:11:1, Rebhan:10:1}. Depending on the holographic implementation of the chiral chemical potential $\mu_A$ and on the kinematical regimes in which the anomalous transport coefficients were defined, different answers for $\sigma_{CME}$ were obtained, ranging from the conventional values (\ref{CSEClassicExpression}) and (\ref{CMEClassicExpression}) to zero. In particular, in \cite{Rubakov:10:1,Rebhan:10:1} it was argued on a very general grounds that $\sigma_{CME}$ should vanish in a gauge theory with gauge invariant UV regularization. Explicit calculation for free continuum fermions with a Pauli-Villars regulator confirms this conclusion \cite{Ren:11:1}. However, it was also argued that if one couples the chiral chemical potential to the conserved axial charge (which differs from the non-conserved axial charge $q_R - q_L$ by a Chern-Simons term $\frac{1}{4 \pi^2} \epsilon_{ijk} A_i \partial_j A_k$), the conventional result (\ref{CMEClassicExpression}) is reproduced \cite{Rubakov:10:1}. It turns out to be completely saturated by the Chern-Simons term. In \cite{Landsteiner:11:2, Landsteiner:12:1, Landsteiner:13:2, Rebhan:10:1, Gynther:10:1} it was shown that different expressions for the chiral magnetic conductivity obtained in holographic models and in hydrodynamical/kinetic theory approximations in fact correspond to different definitions of the axial and vector currents, the ``consistent'' and the ``covariant'' ones.

 A detailed analysis performed in \cite{Kharzeev:09:1, Ren:11:1} (and later in \cite{Burkov:13:1} with application to Weyl semimetals) for free fermions showed that the limit of zero momentum and zero frequency which one should take in order to calculate the chiral magnetic conductivity (\ref{CMEClassicExpression}) is highly singular and should be carefully regularized. More recently it was also shown that the definition of $\sigma_{CME}$ for a constant magnetic field in a finite volume corresponds to a singular point of the one-loop QED polarization tensor \cite{Buividovich:13:6}. Again, depending on the order of limits and the definition of observables one can get different results for $\sigma_{CME}$. All these observations might indicate some sort of infrared instability of chirally imbalanced matter in the presence of a dynamical electromagnetic field. Possible infrared instabilities of interacting chiral medium which might affect the Chiral Magnetic Effect were discussed recently in \cite{Yamamoto:13:1, Sadofyev:13:1, Sadofyev:13:2}.

 To summarize, the questions of whether the chiral magnetic and chiral separation conductivities are affected by interactions and how they are related to axial anomaly remain to a large extent unsettled. Existing proofs of exactness work only in certain kinematical limits and (sometimes implicitly) rely on the validity of hydrodynamical approximation \cite{Son:09:1, Sadofyev:10:1, Sadofyev:10:2, Zakharov:12:1}, the finiteness of static correlation length \cite{Jensen:12:1, Jensen:12:2, Banerjee:12:1} or the existence of Fermi surface with well-defined quasiparticle excitations around it \cite{Son:12:2, Stephanov:12:1}. These assumptions appear to be quite nontrivial in the context of strongly interacting quantum field theories, for which, e.g., the non-Fermi-liquid behavior is quite typical \cite{Shankar:94:1, DasSarma:07:1}. Thus fully non-perturbative first-principle studies of anomalous transport properties in lattice gauge theories might be necessary in order to understand the validity of different approximations mentioned above and to measure possible corrections to anomalous transport coefficients. Even in situations where the low-momentum behavior of anomalous current-current correlators does not change due to interactions, it might still be important to measure them for all values of frequency or momentum. Indeed, realistic examples of chirally imbalanced matter such as quark-gluon plasma in heavy ion collisions or quasiparticles in Weyl semi-metals might be quite far from the validity of hydrodynamical/kinetic theory approximations. For such systems the methods of lattice gauge theory might be very helpful.

 As a preparatory step for the numerical study of chiral separation and chiral magnetic effects in lattice gauge theories, in this work we study the anomalous correlators of vector and axial currents for free overlap fermions on the lattice. We pursue several goals: first, we would like to compute the anomalous axial-vector and vector-vector correlators which enter the Kubo formulae for the chiral separation and the chiral magnetic conductivities using a fully self-consistent regularization of chiral fermions. To this end we use overlap Dirac operator coupled to the conventional or chiral chemical potentials \cite{Buividovich:13:6}. The use of the overlap Dirac operator is motivated by the observation that in the standard calculation \cite{Kharzeev:09:1} of the chiral magnetic conductivity the sum over chiral states and the integral over loop momentum do not commute, and thus a careful regularization of individual contributions of different chiral states is required. As we will see, simple summation over chiral states before momentum integration leads to the current-current correlators which do not satisfy the Ward identities. On the other hand, if all the Ward identities are satisfied, the chiral magnetic conductivity should vanish \cite{Rubakov:10:1}. Pauli-Villars regularization of the chiral magnetic conductivity for free massless fermions was considered in \cite{Ren:11:1} and a zero result was obtained. However, Pauli-Villars regularization might not work properly if the left- and right-handed fermions feel different gauge fields (or different chemical potentials) \cite{Neuberger:93:1, Slavnov:93:1}. The reason is that in this case one needs separate left- and right-handed regulator fields, which, however, cannot be given a Dirac mass term. Historically, the solution of this problem was first found in terms of an infinite number of Pauli-Villars regulator fields \cite{Neuberger:93:1, Slavnov:93:1}. A practical development of this construction which is commonly used nowadays is the overlap Dirac operator on the lattice \cite{Neuberger:98:1}. Therefore it seems that overlap fermions are most suitable to address the issue of anomalous transport on the lattice.

 Second, we also clarify the relation of chiral separation and chiral magnetic conductivities and the anomalous current-current correlators to axial anomaly. We show that at least for free chiral fermions this relation holds in the limit when the momentum carried by electromagnetic field quanta is much larger than chemical potentials $\mu_A$ and $\mu_V$. We argue that anomalous current-current correlators might receive corrections in interacting theories due to the existence of the Fermi surface. Thus it would not be unreasonable to study them on the lattice.

 Finally, as a preparation for further lattice measurements of anomalous vector-vector and axial-vector correlators in interacting gauge theories, we would like to estimate finite-volume and finite-spacing artifacts in these observables. We would like also to understand how robust are anomalous transport phenomena against different lattice discretizations of the Dirac operator and the vector and axial current operators.

 The structure of the paper is the following: in Section \ref{sec:definitions} we briefly summarize the definitions of the chiral separation and the chiral magnetic conductivities in terms of anomalous correlators of vector and axial currents \cite{Kharzeev:09:1, Landsteiner:11:2, Landsteiner:12:1} and discuss the relevant kinematical regimes. Then in Section \ref{sec:correlators_continuum} we calculate these correlators in the continuum theory using a finite-temperature regularization (which amounts to integrating out the temporal components of the loop momenta first) and point out that such calculation of chiral magnetic conductivity, while giving an ultraviolet finite answer, might wrongly count chiral states and thus might produce an incorrect result. In contrast, the calculation of the chiral separation conductivity does not contain any hidden ambiguities and yields the conventional value (\ref{CSEClassicExpression}). We also demonstrate that the role of the Dirac mass in the Chiral Separation and the Chiral Magnetic Effects is completely different. While for the former it simply introduces an energy threshold for production of real physical particles, for the latter it smears the Fermi-Dirac distribution and changes its asymptotically exponential decay into a power-law decay. This makes the notion of a distinct Fermi surface quite vague. Some technical details of these calculations are relegated to Appendices \ref{apdx:cse_continuum} and \ref{apdx:cme_continuum}. Following \cite{Ren:11:1}, we then apply the Pauli-Villars regularization and show that with such regularization the chiral magnetic conductivity is indeed equal to zero. We demonstrate that zero result appears due the non-commutativity of the limits of zero spatial momentum and zero chiral chemical potential in the current-current correlator. We also argue that at spatial momentum which is much larger than the chiral chemical potential the anomalous vector-vector correlator should be an asymptotically linear function of momentum with a finite slope which is equal to minus the conventional result (\ref{CMEClassicExpression}) for the chiral magnetic conductivity. On the other hand, if one makes the axial charge conserved by including the Chern-Simons term into its definition \cite{Rubakov:10:1, Ren:11:1}, the conventional value of the chiral magnetic conductivity (\ref{CMEClassicExpression}) is recovered.

 In Section \ref{sec:two_point_overlap} we construct the anomalous current-current correlators and calculate the chiral separation and the chiral magnetic conductivities for free overlap fermions on the lattice. Technical details of the calculations are given in Appendix \ref{apdx:overlap_derivatives}. The results turn out to coincide with the ones obtained in the Pauli-Villars regularization up to some finite-volume effects. We also find that the same results are reproduced even for the massless Wilson-Dirac fermions for which chiral symmetry is realized only at low energies. On the other hand, measurements with non-conserved vector currents reproduce the naive result (\ref{CMEClassicExpression}) which is an artefact of a wrong counting of chiral states. In addition, we demonstrate that for overlap fermions nontrivial realization of lattice chiral symmetry in terms of L\"{u}scher transformations \cite{Luscher:98:1} is essential to reproduce anomalous transport phenomena.

 In Section \ref{sec:rel_to_anomaly} we clarify the relation of chiral separation and chiral magnetic conductivities to axial anomaly. We argue that vector and axial Ward identities fix the behavior of the anomalous current-current correlators at large spatial momenta (much larger than the corresponding chemical potentials). Namely, the vector-axial correlator at finite chemical potential vanishes at large momenta, and the vector-vector correlator at finite chiral chemical potential has asymptotically linear behavior. We demonstrate that in order to accurately define the coefficient of this linear dependence, one should allow for chiral chemical potential which slowly varies either in time or in space. In both cases this coefficient is equal to minus the conventional result (\ref{CMEClassicExpression}). In the case of time-dependent chiral chemical potential, this value indeed turns out to be directly related to the anomaly coefficient and thus does not receive neither perturbative nor nonperturbative corrections. Following the work \cite{Knecht:04:1} on the perturbative non-renormalization of the transverse parts of the correlator of two vector and one axial currents, we then argue that in the case when the chiral chemical potential varies infinitely slowly in space but is constant in time, the relation of the vector-vector correlator at finite chiral chemical potential to the anomaly holds only perturbatively and non-perturbative corrections might still appear when the chiral symmetry is broken. The limit of time-independent chiral chemical potential appears to be much more natural from the point of view of lattice simulations in Euclidean space. Finally, in the concluding Section \ref{sec:conclusions} we discuss the implications of our results and possible directions of further work.

\section{Anomalous transport coefficients and current-current correlators}
\label{sec:definitions}

 Within the linear response approximation, chiral separation conductivity $\sigma_{CSE}$ can be extracted from the off-diagonal spatial components of the correlator of a vector and an axial currents $j^V_i\lr{x}$ and $j^A_i\lr{x}$ \cite{Son:06:2} at nonzero chemical potential $\mu_V$:
\begin{eqnarray}
\label{CSEPolarizationTensor}
 \Pi^{AV}_{ij}\lr{k} = \int d^4x \, e^{i k_{\mu} x_{\mu}} \, \vev{ j^{A}_i\lr{x} j^{V}_j\lr{0} }_{\mu_V} .
\end{eqnarray}
Similarly, chiral magnetic conductivity $\sigma_{CME}$ can be expressed in terms of the off-diagonal spatial components of the correlator of two vector currents at nonzero chiral chemical potential \cite{Kharzeev:09:1} $\mu_A$:
\begin{eqnarray}
\label{CMEPolarizationTensor}
 \Pi^{VV}_{ij}\lr{k} = \int d^4 x e^{i k_{\mu} x_{\mu}} \vev{j^{V}_i\lr{x} j^{V}_j\lr{0}}_{\mu_A}
\end{eqnarray}
Throughout the paper, we assume that the Greek indices $\mu, \nu, \ldots = 0, 1, 2, 3$ label the components of the vectors and tensors in four-dimensional space and Latin indices $i, j, k, \ldots = 1, 2, 3$ label the corresponding spatial components. We also use the notation $\vec{x}$ to denote vectors in three spatial dimensions.

 As discussed in \cite{Landsteiner:11:2, Landsteiner:12:1}, in order to obtain the anomalous transport coefficients $\sigma_{CSE}$ and $\sigma_{CME}$ which are relevant for hydrodynamical equations, one has first to take the limit of zero frequency and then the limit of zero spatial momentum. Such order of limits has to be contrasted with the one which is relevant for the conventional dissipative transport - first the spatial momentum is sent to zero and then the frequency. In other words, in the case of anomalous transport we study the response of static quantities to static perturbations, for example, the response of spatial electric (or axial) current to the static magnetic field.

 For definiteness, let us assume that only the component $B_1$ of the magnetic field is nonzero and is spatially modulated in the direction $3$: $B_1\lr{x^3} \sim B_1 e^{i k_3 x_3}$. We can work in the gauge in which only the component $V_2$ of the vector gauge field is different from zero, so that $B_1 = -i \, k_3 \, V_2$. According to (\ref{CSEClassicExpression}) and (\ref{CMEClassicExpression}), anomalous transport coefficients $\sigma_{CSE}$ and $\sigma_{CME}$ describe the linear response of the $j_1$ component of the current to the magnetic field $B_1$ and thus can be obtained from the following Kubo formulae \cite{Landsteiner:11:2, Landsteiner:12:1}:
\begin{eqnarray}
\label{Kubo_CSE}
 \sigma_{CSE} = \lim\limits_{k_3 \rightarrow 0}\lim\limits_{k_0 \rightarrow 0} \frac{i}{k_3} \Pi_{12}^{AV}\lr{k_3}  ,
\\
\label{Kubo_CME}
 \sigma_{CME} = \lim\limits_{k_3 \rightarrow 0}\lim\limits_{k_0 \rightarrow 0} \frac{i}{k_3} \Pi_{12}^{VV}\lr{k_3}  .
\end{eqnarray}
We conclude that in order to reproduce the conventional values (\ref{CSEClassicExpression}) and (\ref{CMEClassicExpression}), the current-current correlators (\ref{CSEPolarizationTensor}) and (\ref{CMEPolarizationTensor}) should take the form
\begin{eqnarray}
\label{CSEClassicPolTens}
 \Pi_{12}^{AV}\lr{k_3} = -\frac{i \mu_V k_3}{2 \pi^2}   ,
\\
\label{CMEClassicPolTens}
 \Pi_{12}^{VV}\lr{k_3} = -\frac{i \mu_A k_3}{2 \pi^2}  ,
\end{eqnarray}
at small spatial momentum $k_3$.

\section{Anomalous current-current correlators for free fermions in the continuum}
\label{sec:correlators_continuum}

\subsection{Chiral Separation Effect}
\label{subsec:cse_continuum}

 It is instructive to consider first the case of the Chiral Separation Effect, since the chiral separation conductivity turns out to be completely free from ultraviolet divergences and one can analytically obtain an unambiguous answer for the correlator of vector and axial currents (\ref{CSEPolarizationTensor}) at nonzero spatial momentum. The results which we obtain in this Subsection will be also necessary for the discussion of the Chiral Magnetic Effect in the next Subsection \ref{subsec:cme_continuum}. For further comparison with the calculation of the chiral magnetic conductivity we introduce a generic Dirac mass $m$. As we will see, the limit $m \rightarrow 0$ turns out to be straightforward.

 For free fermions, the polarization tensor (\ref{CSEPolarizationTensor}) is given by
\begin{widetext}
\begin{eqnarray}
\label{CSEPolTens1}
 \Pi^{AV}_{\mu\nu}\lr{k} = \int \frac{d^4 l}{\lr{2 \pi}^4}
 \tr\lr{\gamma_{\mu} \gamma_5 \mathcal{D}^{-1}\lr{l + k/2, \mu_V} \gamma_{\nu} \mathcal{D}^{-1}\lr{l - k/2, \mu_V}}
 = \nonumber \\ =
 \int \frac{d^4 l}{\lr{2 \pi}^4}
 \frac{\tr\lr{\gamma_{\mu} \gamma_5 \lr{m - i \gamma_{\alpha}\lr{l_{\alpha} + k_{\alpha}/2}} \gamma_{\nu} \lr{m - i \gamma_{\beta}\lr{l_{\beta} - k_{\beta}/2}}}}{\lr{\lr{l+k/2}^2 + m^2} \lr{\lr{l - k/2}^2 + m^2}} ,
\end{eqnarray}
\end{widetext}
where $\mathcal{D}\lr{p, \mu_V} = i \gamma_{\mu} p_{\mu} + \mu_V \gamma_0 + m$ is the massive Euclidean Dirac operator at finite chemical potential $\mu_V$. In the second line of (\ref{CSEPolTens1}) we have to shift the contour of integration over the time-like component of the loop momentum $l_0$ from the real axis to the axis $\Im l_0 = -\mu_V$: $l_0 \rightarrow l_0 - i \mu_V$. For simplicity, in this work we consider only the limit of zero temperature.

\begin{figure}
  \centering
  \includegraphics[width=6cm]{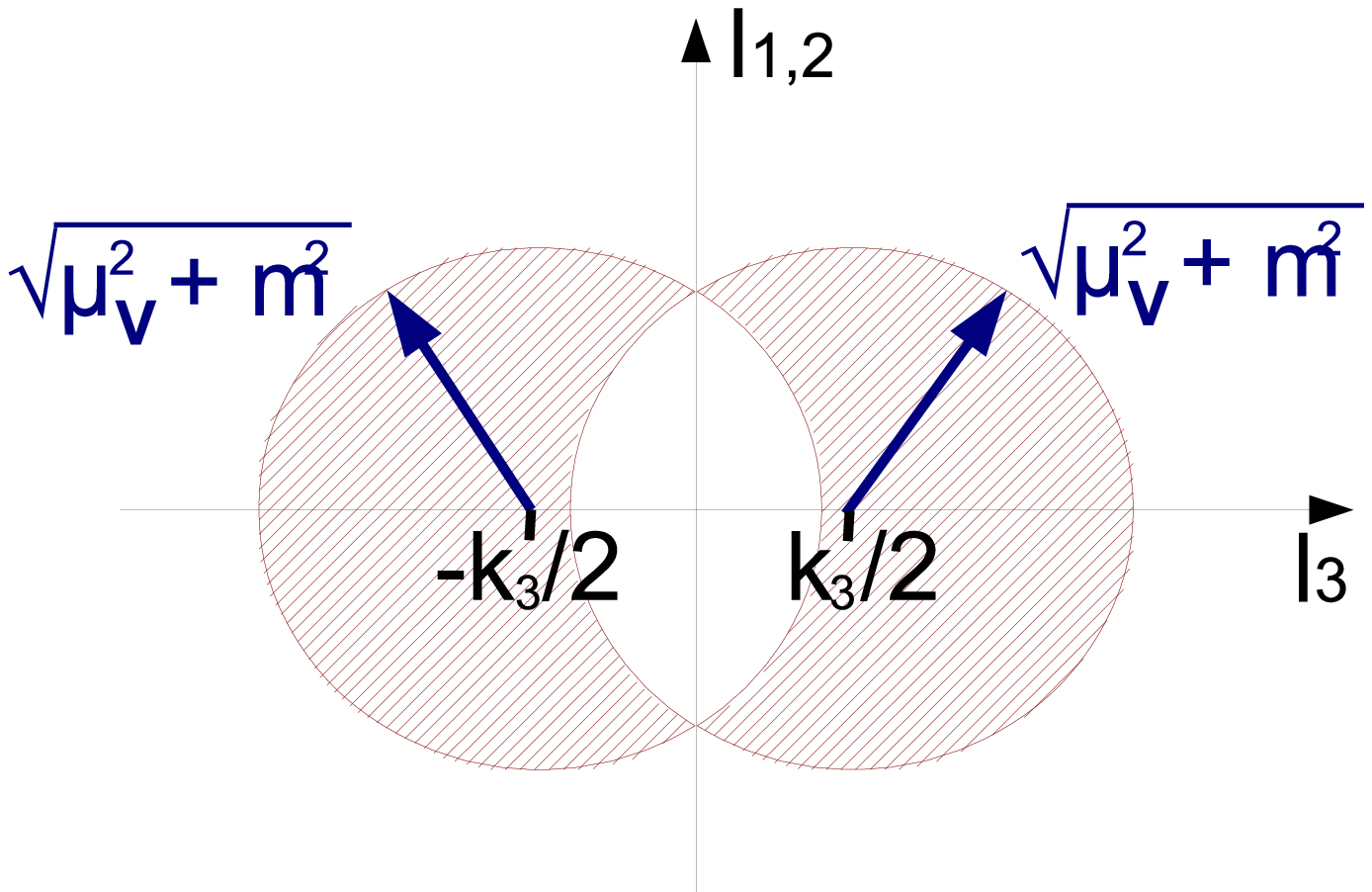}\\
  \includegraphics[width=6cm]{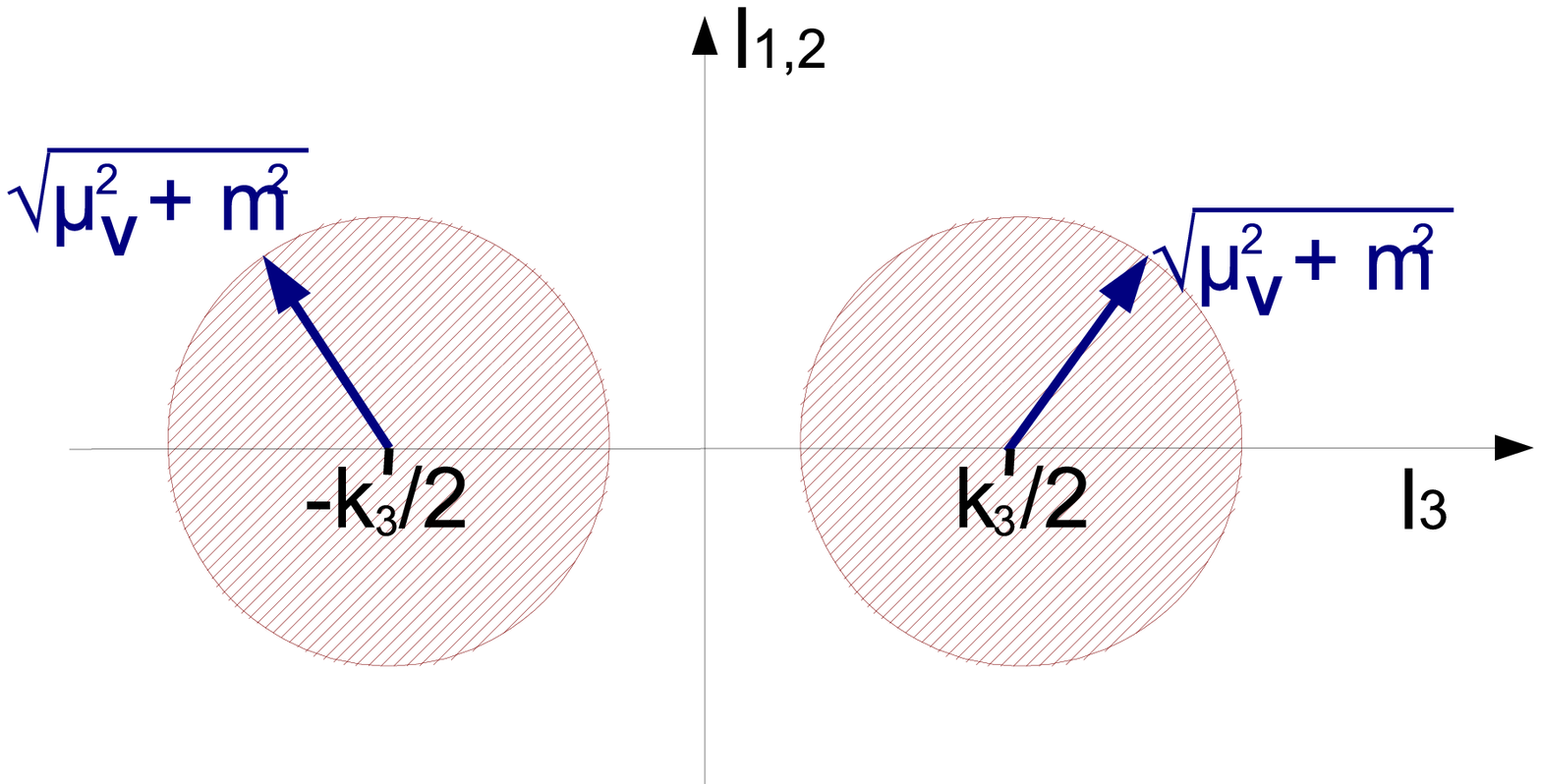}\\
  \caption{Regions in the momentum space which contribute to the integral over spatial loop momentum in (\ref{CSEPolTens2}). Above: for external momentum $k_3 < 2 \mu_V$, below: for $k_3 > 2 \mu_V$.}
  \label{fig:finite_mu_regions}
\end{figure}

 In order to calculate (\ref{CSEPolTens1}), we use the technique of finite-temperature Euclidean quantum field theory and first integrate out the (Euclidean) time-like component of the loop momentum $l_0$ (cf. \cite{Kharzeev:09:1, Ren:11:1}). The details of this calculation are given in Appendix \ref{apdx:cse_continuum}. For the component $\Pi_{12}^{AV}\lr{k_3}$ of the current-current correlator (\ref{CSEPolarizationTensor}) which enters the Kubo relations (\ref{Kubo_CSE}) we obtain at positive values of chemical potential $\mu_V > m$:
\begin{eqnarray}
\label{CSEPolTens2}
 \Pi^{AV}_{12}\lr{k_3} = i \int\limits_{-\infty}^{+\infty} \frac{d l_3}{2 \pi l_3} \int\limits_{0}^{+\infty}
  \frac{d\lr{\pi l_{\perp}^2}}{\lr{2 \pi}^2}
\nonumber \\
  \left(
 \theta\lr{\mu_V - \sqrt{\lr{l_3 + k_3/2}^2 + l_{\perp}^2 + m^2}}
 - \right. \nonumber \\ \left. -
 \theta\lr{\mu_V - \sqrt{\lr{l_3 - k_3/2}^2 + l_{\perp}^2 + m^2}}
 \right) ,
\end{eqnarray}
where we have denoted $l_{\perp} = \sqrt{l_1^2 + l_2^2}$. From this expression we see that the integration over spatial loop momentum is restricted to two disjoint regions in momentum space which are depicted on Fig.~\ref{fig:finite_mu_regions}. If the external momentum $|\vec{k}| > 2 \sqrt{\mu_V^2 - m^2}$, these regions are just two spheres of radii $\sqrt{\mu_V^2 - m^2}$ centered around $l_3 = \pm k_3/2$, $l_{1,2} = 0$. For $|k_3| < 2 \sqrt{\mu_V^2 - m^2}$, the intersection of these two spheres is removed from the integral. In particular, in the limit $|k_3| \ll \sqrt{\mu_V^2 - m^2}$ the integration region is just a thin shell around the Fermi surface $|\vec{l}| = \mu_V$. The integrand is even in $l_3$ and thus the two regions at $l_3 > 0$ and at $l_3 < 0$ yield equal contributions to the integral.

\begin{figure}
  \centering
  \includegraphics[width=6cm, angle=-90]{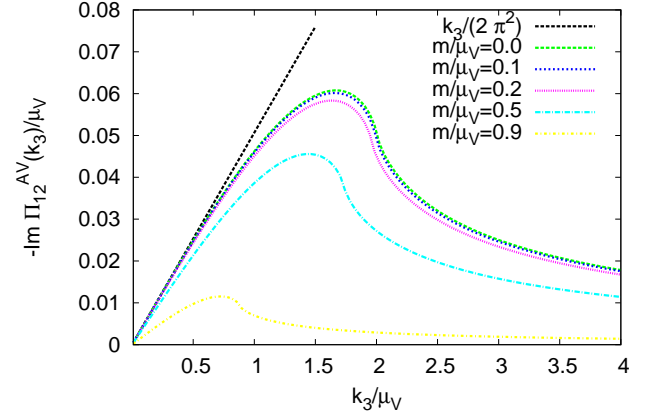}\\
  \caption{The correlator of axial and vector currents $\Pi_{12}^{AV}\lr{k_3}$ calculated for free Dirac fermions (see expressions (\ref{CSEPolTens})) as a function of spatial momentum $k_3$ for different values of the Dirac mass $m$. Black dashed line corresponds to the linear dependence $\Pi_{12}\lr{k_3} = -\frac{i k_3 \mu}{2 \pi^2}$ which reproduces the expression (\ref{CSEClassicExpression}).}
  \label{fig:cse_theory}
\end{figure}

 The integration over spatial loop momentum in (\ref{CSEPolTens2}) can be performed analytically, and we finally obtain the closed expression for $\Pi^{AV}_{12}\lr{k_3}$:
\begin{eqnarray}
\label{CSEPolTens}
 \Pi^{AV}_{12}\lr{k_3} =
 -\frac{i}{\lr{2 \pi}^2}
 \left( \sqrt{\mu_V^2 - m^2} k_3
 - \right. \nonumber \\ \left. -
 \lr{\mu_V^2 - m^2 - k_3^2/4} \, \log\absval{\frac{2 \sqrt{\mu_V^2 - m^2} - k_3}{2 \sqrt{\mu_V^2 - m^2} + k_3}}
 \right) .
\end{eqnarray}
This expression is of course valid only for $\mu_V > m$. For $\mu_V < m$ $\Pi^{AV}_{12}\lr{k_3}$ is equal to zero at any value of $k_3$. The function $\Pi^{AV}_{12}\lr{k_3}$ given by the above expression is plotted on Fig.~\ref{fig:cse_theory} for different values of the Dirac mass $m$. At momentum $k_3$ smaller than approximately $1.7 \, \sqrt{\mu_V^2 - m^2}$ it grows linearly with $k_3$, with the slope being equal to the conventional expression (\ref{CSEClassicExpression}). Indeed, expanding the logarithm in (\ref{CSEPolTens}) in powers of $k_3$ we get
\begin{eqnarray}
\label{CSEPolTensSmallK}
 \Pi^{AV}_{12}\lr{k_3} =
 -\frac{i k_3 \sqrt{\mu_V^2 - m^2}}{2 \pi^2}
 + \nonumber \\ +
 \frac{i k_3^3}{24 \pi^2 \sqrt{\mu_V^2 - m^2}} + O\lr{k_3^5} .
\end{eqnarray}
Inserting now the expression (\ref{CSEPolTens}) into the Kubo formula (\ref{Kubo_CSE}) and taking into account the expansion (\ref{CSEPolTensSmallK}), we conclude that at zero mass the expression (\ref{CSEPolTens}) reproduces the conventional result (\ref{CSEClassicExpression}).

 At $k_3 = 2 \sqrt{\mu_V^2 - m^2}$ the function $\Pi^{AV}_{12}\lr{k_3}$ has a singularity at which the derivative over $k_3$ diverges. This singularity corresponds to the transition from the regime of small $k_3$, in which the correlator (\ref{CSEPolTens2}) is saturated by an integral over a shell surrounding the Fermi surface, to the regime of large $k_3$, for which the integration over loop momentum goes over two widely separated disjoint spheres in momentum space. Physically, the transition between the two regimes can be interpreted in the following way: at small $k_3$ virtual photon with momentum $k_3$ can kick out only pairs of fermions which live very close to the Fermi surface (and have nearly opposite momenta), and thus the volume of phase space available for such process is limited. As $k_3$ grows, this volume and hence the process amplitude gradually increase until at $k_3 = 2 \mu_V$ it becomes possible to kick out fermions from everywhere within Fermi surface. Then, however, the suppressing factor $1/l_3$ in (\ref{CSEPolTens2}) enters the game, and the amplitude again starts decreasing. Indeed, in the limit of large spatial momentum $k_3 \rightarrow +\infty$, the values of $l_3$ which contribute to the integral in (\ref{CSEPolTens}) are very close to $\pm k_3/2$, and the correlator (\ref{CSEPolTens}) goes to zero as
\begin{eqnarray}
\label{CSEPolTensLargeK}
 \Pi^{AV}_{12}\lr{k_3} =
 -\frac{2 i \lr{\mu_V^2 - m^2}^{3/2}}{3 \pi^2 k_3} + O\lr{1/k_3^3} .
\end{eqnarray}

 It is also straightforward to understand the role played by the Dirac mass $m$: it simply introduces a threshold below which the chemical potential $\mu_V$ cannot excite any fermionic state from the vacuum, so that the chiral separation effect is absent for $\mu_V < m$. We can also interpret this fact as the saturation of the Chiral Separation Effect by physical particle states which are only available at nonzero chemical potential $\mu_V > m$.

\subsection{Chiral Magnetic Effect from the two-point function}
\label{subsec:cme_continuum}

 In this Subsection we calculate the chiral magnetic conductivity $\sigma_{CME}$ for free continuum fermions at finite chiral chemical potential $\mu_A$. Our aim here is to illustrate a subtle point in the previous calculations which is related to the ambiguity in the counting of chiral states. To this end we essentially repeat the derivations of \cite{Kharzeev:09:1, Ren:11:1, Son:13:1} but treat the chirality more carefully.

 For free fermions the current-current correlator (\ref{CMEPolTens}) which enters the Kubo relations (\ref{Kubo_CME}) is given by
\begin{eqnarray}
\label{CMEPolTens1}
 \Pi^{VV}_{\mu\nu}\lr{k} = \int \frac{d^4 l}{\lr{2 \pi}^4}
 \nonumber \\
 \tr\lr{\gamma_{\mu} \mathcal{D}^{-1}\lr{l + k/2, \mu_A} \gamma_{\nu} \mathcal{D}^{-1}\lr{l - k/2, \mu_A}} ,
\end{eqnarray}
where $\mathcal{D}\lr{p, \mu_A} = i \gamma_{\mu} p_{\mu} + \mu_A \gamma_0 \gamma_5 + m$ is now the massive Dirac operator with chiral chemical potential $\mu_A$. The Dirac propagator $\mathcal{D}^{-1}\lr{p, \mu_A}$ can be represented in the following chiral block form (see Appendix \ref{apdx:cme_continuum} and also \cite{Kharzeev:09:1, Ren:11:1} for more details):
\begin{eqnarray}
\label{propagator_mu5_chir_decomp}
 \mathcal{D}^{-1}\lr{p, \mu_A}
  =
 \sum\limits_{s = \pm} G_s\lr{p, \mu_A} \mathcal{P}_s
 \times \nonumber \\ \times
 \left(
   \begin{array}{cc}
                                  m & -i p_0 + \mu_A - s |\vec{p}| \\
       -i p_0 - \mu_A + s |\vec{p}| &                            m \\
   \end{array}
 \right)  ,
\end{eqnarray}
where $G_s\lr{p, \mu_A} = {p_0^2 + \lr{|\vec{p}| - s \mu_A}^2 + m^2}$, $\dslash{p} = \sigma_k p_k$, $p_k$ are spatial components of the momentum and $\sigma_k$ with $k = 1 \ldots 3$ are the Pauli matrices. The operators $\mathcal{P}_{\pm} = \frac{1 \pm \dslash{p}/|\vec{p}|}{2}$ are the projectors of spin states on the direction of momentum and thus project out states of definite chirality. Correspondingly, summands with $s = \pm$ in (\ref{propagator_mu5_chir_decomp}) are the propagators of states with definite chirality.

 We now insert the decomposition (\ref{propagator_mu5_chir_decomp}) into (\ref{CMEPolTens1}) and again perform integration over the time-like component of the loop momentum $l_0$. Details of this calculation are summarized in Appendix \ref{apdx:cme_continuum}. The result is
\begin{eqnarray}
\label{CMEPolTens2}
 \Pi^{VV}_{12}\lr{k_3} = -i \int \frac{d^3 l}{\lr{2 \pi}^3}
 \sum\limits_{s,s'} \, \frac{s q p_3 - s' p q_3}{2 p q \lr{p - s' s q}}
 \times \nonumber \\ \times
 \lr{
 \frac{p - s \mu_A}{\sqrt{\lr{p - s \mu_A}^2 + m^2}}
 -
 \frac{s s' \lr{q - s' \mu_A}}{\sqrt{\lr{q - s' \mu_A}^2 + m^2}}
 }  ,
\end{eqnarray}
where $p = |\vec{l} + \vec{k}/2|$ and $q = |\vec{l} - \vec{k}/2|$ denote the absolute values of momenta of two virtual fermions saturating the integral in (\ref{CMEPolTens1}) and $s$, $s'$ denote their chiralities.

 From this representation we see immediately that the individual contributions of states with different chiralities $s$ and $s'$ to the current-current correlator (\ref{CMEPolTens2}) are in fact divergent. Let us try, however, to sum up all of these contributions before performing the integration over spatial loop momentum $\vec{l}$, as it is done, for example, in \cite{Kharzeev:09:1}. After some algebraic manipulations which are given in Appendix \ref{apdx:cme_continuum} we arrive at
\begin{eqnarray}
\label{CMEPolTens3}
 \Pi^{VV}_{12}\lr{k_3} = -i \int \frac{d^3 l}{\lr{2 \pi}^3}
 \frac{1}{2 l_3}
 \times \nonumber \\ \times
 \left(
 \frac{p - \mu_A}{\sqrt{\lr{p - \mu_A}^2 + m^2}}
 -
 \frac{q - \mu_A}{\sqrt{\lr{q - \mu_A}^2 + m^2}}
 + \right. \nonumber \\ \left. +
 \frac{q + \mu_A}{\sqrt{\lr{q + \mu_A}^2 + m^2}}
 -
 \frac{p + \mu_A}{\sqrt{\lr{p + \mu_A}^2 + m^2}}
 \right)  ,
\end{eqnarray}
where different summands in the brackets are the contributions of the states with different chiralities.

 At this point it is interesting to see how different is the role of the Dirac mass in the Chiral Separation Effect and the Chiral Magnetic Effect. In the case of Chiral Separation Effect the mass played the role of the threshold below which there were no physical particles and no transport occurred. In contrast, in the case of the Chiral Magnetic Effect the mass simply broadens the Fermi-Dirac distribution, which is just a step function at zero mass and in the limit of zero temperature. The physical origin of such broadening is clear: Dirac mass induces transitions between different chirality sectors which excite chiral states with momenta even outside of the Fermi surface. While at zero mass and nonzero temperature the density of particles in momentum space is exponentially suppressed outside of the Fermi surface, at nonzero mass this exponential suppression turns into a power-law decay. Such power-law decay of particle density in momentum space might lead to some interesting non-Fermi-liquid behavior when the interactions are switched on.

\begin{figure}
  \centering
  \includegraphics[width=6cm, angle=-90]{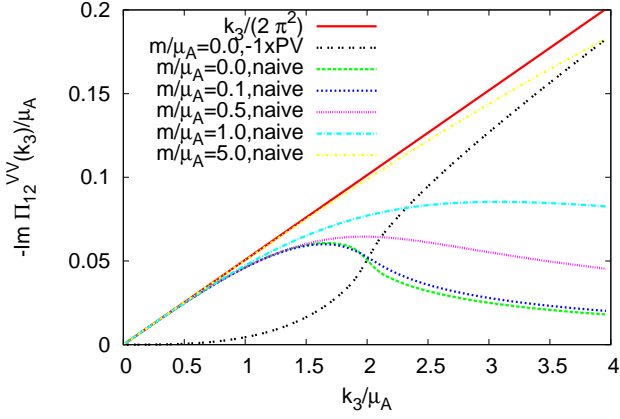}\\
  \caption{The correlator of two vector currents $\Pi_{12}^{VV}\lr{k_3}$ calculated for free Dirac fermions (see expression (\ref{CMEPolTens})) as a function of spatial momentum $k_3$ for different values of the Dirac mass $m$ without any regulator (label ``naive'') and with the Pauli-Villars regularization (label ``PV''). Red solid line corresponds to the linear dependence $\Pi_{12}^{VV}\lr{k_3} = -\frac{i k_3 \mu_A}{2 \pi^2}$ which reproduces the expression (\ref{CMEClassicExpression}).}
  \label{fig:cme_theory}
\end{figure}

 Let us consider first the limit of zero mass and assume, for definiteness, that $\mu_A > 0$. The fractions in the second line then turn into the sign functions, and after some simple manipulations we get
\begin{eqnarray}
\label{CMEPolTensMassless}
 \Pi^{VV}_{12}\lr{k_3}
 = \nonumber \\ =
 -i \int \frac{d^3 l}{\lr{2 \pi}^3} \frac{1}{l_3} \, \lr{\theta\lr{p - \mu_A} - \theta\lr{q - \mu_A}}
 = \nonumber \\ =
 i \int \frac{d^3 l}{\lr{2 \pi}^3} \frac{1}{l_3} \, \lr{\theta\lr{\mu_A - p} - \theta\lr{\mu_A - q}}  ,
\end{eqnarray}
that is, exactly the massless limit of the expression (\ref{CSEPolTens2}) for the correlator of axial and vector currents with the replacement $\mu_V \rightarrow \mu_A$. We see that naively chiral magnetic conductivity turns out to be exactly equal to the conventional value (\ref{CMEClassicExpression}), in complete agreement with the results of \cite{Kharzeev:09:1}. Again, note a subtle point in the transformation from the expression (\ref{CMEPolTens3}) to (\ref{CMEPolTensMassless}) and then from the second to the third line of (\ref{CMEPolTensMassless}): individual contributions of different chiralities involve factors $\theta\lr{p - \mu_A}$ and $\theta\lr{q - \mu_A}$ which cut out a non-compact region in the momentum space and lead to the non-convergent integral over spatial loop momentum. Localization of integrand in the compact regions (see Fig.~\ref{fig:finite_mu_regions}) only occurs after summing up the contributions of all chiralities. We conclude that the trace over states of different chiralities and the spatial momentum integral do not commute. This again suggests that the polarization tensor (\ref{CMEPolarizationTensor}) should be carefully regularized. In what follows we will apply the lattice regularization with overlap fermions, which is one of the few fully consistent regularizations of chiral fermions \cite{Slavnov:93:1, Neuberger:93:1}. However, in order to foresee the result, it is instructive to consider the simplest Pauli-Villars regularization, as it was done in \cite{Ren:11:1}. Although the Pauli-Villars regularization is rather crude and might not properly take into account the chiral properties, we will see that for our purposes it works quite well and yields the result which agrees with the one obtained using overlap fermions.

 To this end we consider the expression (\ref{CMEPolTens3}) in the limit when the mass $m$ (which would play the role of the Pauli-Villars regulator) is much larger than the external momentum $k_3$ and the chiral chemical potential $\mu_A$. A simple calculation which is given in Appendix \ref{apdx:cme_continuum} shows that for infinitely heavy fermions the linear behavior of $\Pi_{12}^{VV}\lr{k_3}$ both in momentum $k_3$ and in the chiral chemical potential $\mu_A$ persists for all values of $k_3$ and $\mu_A$ which are much smaller than $m$ (cf. \cite{Ren:11:1}):
\begin{eqnarray}
\label{CMEPolTensHeavy}
 \lim\limits_{m \rightarrow \infty} \Pi^{VV}_{12}\lr{k_3}
 = - \frac{i \mu_A k_3}{2 \pi^2}  .
\end{eqnarray}
Introducing now infinitely heavy regulator fermions and subtracting their contribution (\ref{CMEPolTensHeavy}) from the correlator $\Pi^{VV}_{12}\lr{k_3}$ in the massless limit (see expression (\ref{CMEPolTensMassless})), we obtain for the regularized correlator $\tilde{\Pi}^{VV}_{12}\lr{k_3}$:
\begin{eqnarray}
\label{CMEPolTens}
 \tilde{\Pi}^{VV}_{12}\lr{k_3}
 = \nonumber \\ =
 \frac{i}{\lr{2 \pi}^2}
 \lr{ \mu_A k_3
 + \lr{\mu_A^2 - k_3^2/4} \, \log\absval{\frac{2 \mu_A - k_3}{2 \mu_A + k_3}} }
\end{eqnarray}
This correlator is plotted on Fig.~\ref{fig:cme_theory} as a function of spatial momentum $k_3$. For illustration, we also plot the naive result which can be obtained from (\ref{CMEPolTens3}) by naively summing the contributions of different chiralities inside the momentum integral. One can see that as the Dirac mass $m$ increases, the naive result tends to the simple linear dependence (\ref{CMEClassicPolTens}) for larger and larger momenta. Correspondingly, at small $k_3$ the regularized vector-vector correlator $\tilde{\Pi}^{VV}_{12}\lr{k_3}$ behaves as $k_3^3$. The behavior at small $k_3$ can be understood from the expansion (\ref{CSEPolTensSmallK}) of the axial-vector correlator $\Pi^{AV}_{12}\lr{k_3}$ which completely coincides with the unregularized vector-vector correlator $\Pi^{VV}_{12}\lr{k_3}$ upon the replacement $\mu_V \rightarrow \mu_A$. The contribution (\ref{CMEPolTensHeavy}) of Pauli-Villars fermions removes the linear term in the expansion (\ref{CSEPolTensSmallK}), but leaves higher nonlinear terms. From Kubo formulae (\ref{Kubo_CME}) we thus conclude that after the integration over the states of different chiralities in (\ref{CMEPolTens3}) is regularized, the chiral magnetic conductivity turns out to be zero. Since the Pauli-Villars regularization respects the vector gauge invariance, this conclusion is in agreement with the statement of \cite{Rubakov:10:1} that the chiral magnetic conductivity has to vanish in any gauge-invariant regularization. On the other hand, at large momenta the regulator contribution to (\ref{CMEPolTens}) becomes dominant, and thus at $k_3 \gg 2 \mu_A$ the regularized vector-vector correlator asymptotically approaches the linear behavior $\tilde{\Pi}^{VV}_{12}\lr{k_3} = \frac{i \mu_A k_3}{2 \pi^2}$. Taking into account the dependence of the non-regularized expression on the fermion mass (see Fig.~\ref{fig:cme_theory}), we see that as the mass increases, this linear behavior sets in at larger and larger values of $k_3$. This implies that the regularized chiral magnetic conductivity vanishes in the limit of infinite mass, as one could expect from general physical considerations. It is important to note that the coefficient of asymptotically linear dependence is exactly equal to minus the slope of the conventional naive result (\ref{CMEClassicPolTens}) in the limit of small momenta. In Section \ref{sec:rel_to_anomaly} we will see that this coefficient is in fact related to axial anomaly. It is also worth noting that if one includes the Chern-Simons term into the definition of the axial charge in order to make it conserved \cite{Rubakov:10:1, Ren:11:1}, the contribution of this term exactly cancels the contribution of regulator fermions, and the conventional value of the chiral magnetic conductivity (\ref{CMEClassicExpression}) is recovered. The dependence of the chiral magnetic conductivity on the definition of the chiral chemical potential has been also discussed in detail in \cite{Landsteiner:13:2}.

\section{Anomalous current-current correlators for overlap fermions}
\label{sec:two_point_overlap}

 For a generic chiral lattice Dirac operator $\mathcal{D}$, the correlators of axial and vector currents $\vev{j^A_{x, \mu} j^V_{y, \nu}}$ and $\vev{j^V_{x, \mu} j^V_{y, \nu}}$ can be related to the derivatives of the partition function
\begin{eqnarray}
\label{partition_function_def}
 \mathcal{Z} = \int dg_{x, \mu} \det{\mathcal{D}\lrs{g_{x, \mu}, A_{x, \mu}, V_{x, \mu}}}^{N_f} e^{ -S_{YM}\lrs{g_{x, \mu}}}
\end{eqnarray}
over the vector lattice gauge field $V_{x, \mu}$ and the axial lattice gauge field $A_{x, \mu}$ at $V_{x, \mu} = 0$, $A_{x, \mu} = 0$:
\begin{eqnarray}
\label{vv_correlator_def}
 \vev{j^V_{x, \mu} j^V_{y, \nu}}
 =
 \left. T^2 \mathcal{Z}^{-1} \partial^V_{y, \nu} \partial^V_{x, \mu} \mathcal{Z} \right|_{V_{x, \mu} = 0},
 \\
 \label{av_correlator_def}
 \vev{j^A_{x, \mu} j^V_{y, \nu}}
 =
 \left. T^2 \mathcal{Z}^{-1} \partial^V_{y, \nu} \partial^A_{x, \mu} \mathcal{Z} \right|_{V_{x, \mu} = 0, A_{x, \mu} = 0}.
\end{eqnarray}
In the above equations, $T$ is the temperature, $N_f$ is the number of fermion flavors, $g_{x, \mu}$ is the non-Abelian lattice gauge field and $S_{YM}\lrs{g_{x, \mu}}$ is the lattice action of this field. Since we define the axial and the vector currents in terms of the derivatives of the partition function over the corresponding gauge fields, we deal with the ``consistent'' current in the terminology of \cite{Landsteiner:12:1}. To simplify the notation, we have denoted the derivatives over vector and axial lattice gauge fields on the link which goes in the direction $\mu$ from the site $x$ as $\partial^V_{x, \mu} \equiv \frac{\partial}{\partial V_{x, \mu}}$ and $\partial^A_{x, \mu} \equiv \frac{\partial}{\partial A_{x, \mu}}$. In what follows, we will also omit the functional arguments of $\mathcal{D}\lrs{g_{x, \mu}, A_{x, \mu}, V_{x, \mu}}$. In (\ref{vv_correlator_def}) and (\ref{av_correlator_def}) we have also taken into account that for $V_{x, \mu} = 0$ and $A_{x, \mu} = 0$ $\partial^V_{x, \mu} \mathcal{Z} = \mathcal{Z} \vev{j^V_{x, \mu}} = 0$ and $\partial^A_{x, \mu} \mathcal{Z} = \mathcal{Z} \vev{j^V_{x, \mu}} = 0$ by virtue of Lorentz symmetry. We assume that $V_{x, \mu}$ is a non-compact lattice gauge field which enters the fermionic action only in terms of the compact link variables $u_{x, \mu} = e^{i V_{x, \mu}}$. The coupling of the axial gauge field to lattice fermions is quite nontrivial and was considered recently in \cite{Buividovich:13:6, Buividovich:13:7}. However, in order to calculate the correlators (\ref{vv_correlator_def}) and (\ref{av_correlator_def}) we will only need the first derivative of $\mathcal{D}$ over $A_{x, \mu}$ at zero $A_{x, \mu}$, which can be easily expressed in terms of the derivative of $\mathcal{D}$ over $V_{x, \mu}$ \cite{Kikukawa:98:1}. In order to obtain the current-current correlators (\ref{CSEPolarizationTensor}) and (\ref{CMEPolarizationTensor}) in momentum space, we perform the lattice Fourier transform
\begin{eqnarray}
\label{lattice_fourier}
 \Pi^{AV}_{\mu\nu}\lr{k} = \sum\limits_x e^{i k x} \vev{j^A_{x, \mu} j^V_{0, \nu}}  ,
 \nonumber \\
 \Pi^{VV}_{\mu\nu}\lr{k} = \sum\limits_x e^{i k x} \vev{j^V_{x, \mu} j^V_{0, \nu}}  .
\end{eqnarray}

 Direct application of the relations (\ref{vv_correlator_def}) and (\ref{av_correlator_def}) to the partition function (\ref{partition_function_def}) gives:
\begin{eqnarray}
\label{vv_correlator_observable}
 \vev{j^V_{x, \mu} j^V_{y, \nu}}
 =
 -T^2 \, N_f \, \vev{ \tr\lr{\mathcal{D}^{-1} \partial^V_{y, \nu} \mathcal{D} \, \mathcal{D}^{-1} \partial^V_{x, \mu} \mathcal{D} } }
 + \nonumber \\ +
 T^2 \, N_f \, \vev{ \tr\lr{\mathcal{D}^{-1} \partial^V_{y, \nu} \partial^V_{x, \mu} \mathcal{D} } }
 + \nonumber \\ +
 T^2 \, N_f^2 \, \vev{ \tr\lr{\mathcal{D}^{-1} \partial^V_{y, \nu} \mathcal{D}} \tr\lr{\mathcal{D}^{-1} \partial^V_{x, \mu} \mathcal{D}} } ,
 \\
 \label{av_correlator_observable}
 \vev{j^A_{x, \mu} j^V_{y, \nu}}
 =
 -T^2 \, N_f \, \vev{ \tr\lr{\mathcal{D}^{-1} \partial^V_{y, \nu} \mathcal{D} \, \mathcal{D}^{-1} \partial^A_{x, \mu} \mathcal{D} } }
 + \nonumber \\ +
 T^2 \, N_f \, \vev{ \tr\lr{\mathcal{D}^{-1} \partial^V_{y, \nu} \partial^A_{x, \mu} \mathcal{D} } }
 + \nonumber \\ +
 T^2 \, N_f^2 \, \vev{ \tr\lr{\mathcal{D}^{-1} \partial^V_{y, \nu} \mathcal{D}} \tr\lr{\mathcal{D}^{-1} \partial^A_{x, \mu} \mathcal{D}} } ,
\end{eqnarray}
where $\vev{\ldots}$ denotes averaging over non-Abelian gauge field configurations:
\begin{eqnarray}
\label{vev_def}
\vev{ \mathcal{O} } = \int dg_{x, \mu} \mathcal{O}\lrs{g_{x, \mu}}
\det{\mathcal{D}\lrs{g_{x, \mu}}}^{N_f} e^{ -S_{YM}\lrs{g_{x, \mu}}} .
\end{eqnarray}
In (\ref{vv_correlator_observable}) and (\ref{av_correlator_observable}), the first summand is the contribution of a single fermionic loop with two vertices (either two vector vertices or one axial vertex and one vector vertex). The second summand is the contribution of a mixed vector-axial or vector-vector vertex which is present only on the lattice. Finally, the third summand is a contribution of two disconnected fermion loops, each with one vertex. In this work we calculate the correlators (\ref{vv_correlator_def}) and (\ref{av_correlator_def}) for free fermions, therefore we do not perform the averaging (\ref{vev_def}) and simply calculate the expressions (\ref{vv_correlator_observable}) and (\ref{av_correlator_observable}) for free lattice Dirac operator. Obviously, only the connected part of (\ref{vv_correlator_observable}) and (\ref{av_correlator_observable}) are different from zero in this case.

\subsection{Chiral Separation Effect}
\label{subsec:cse_overlap}

\begin{figure*}[Htpb]
\includegraphics[width=5cm, angle=-90]{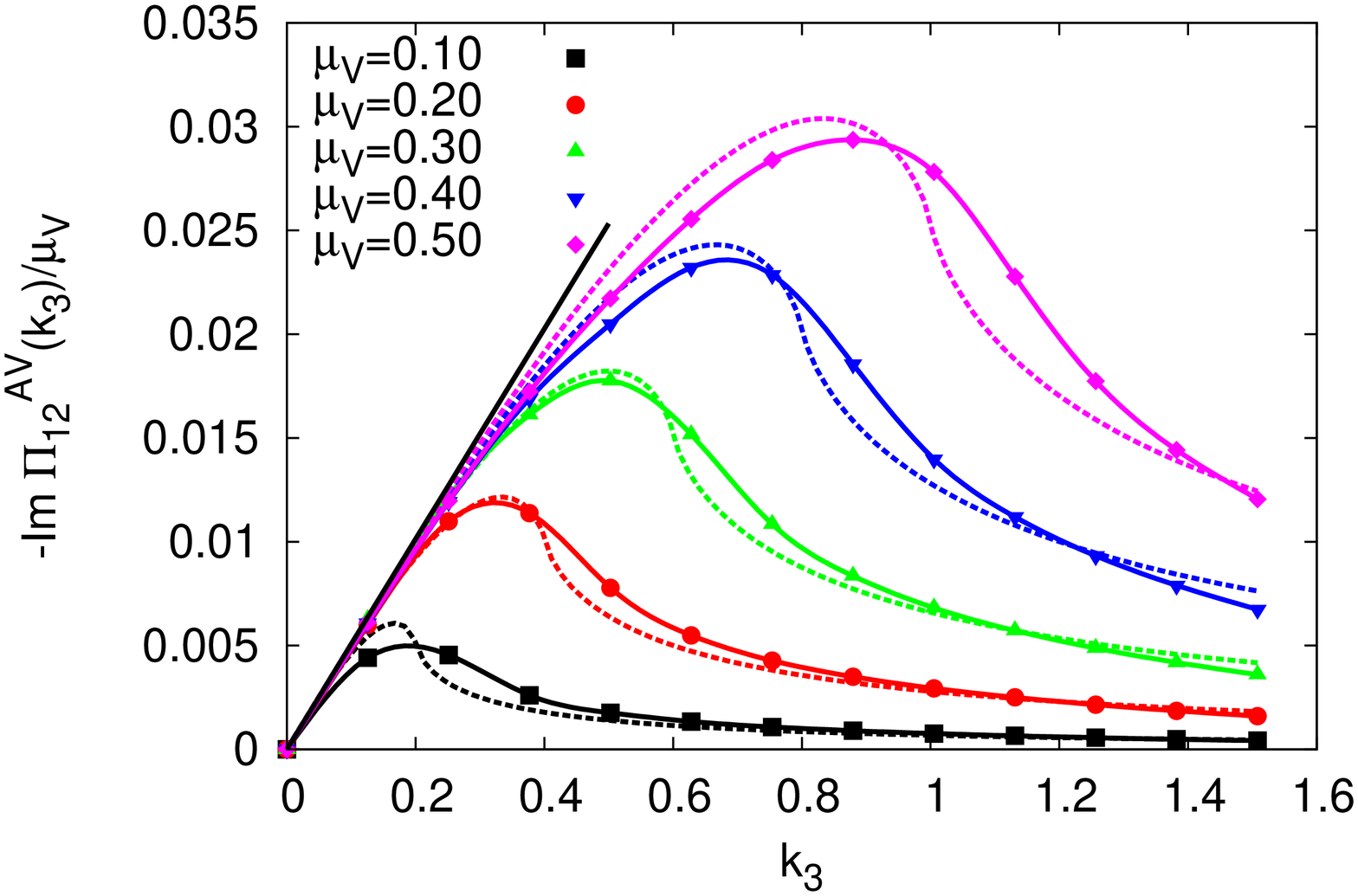}
\includegraphics[width=5cm, angle=-90]{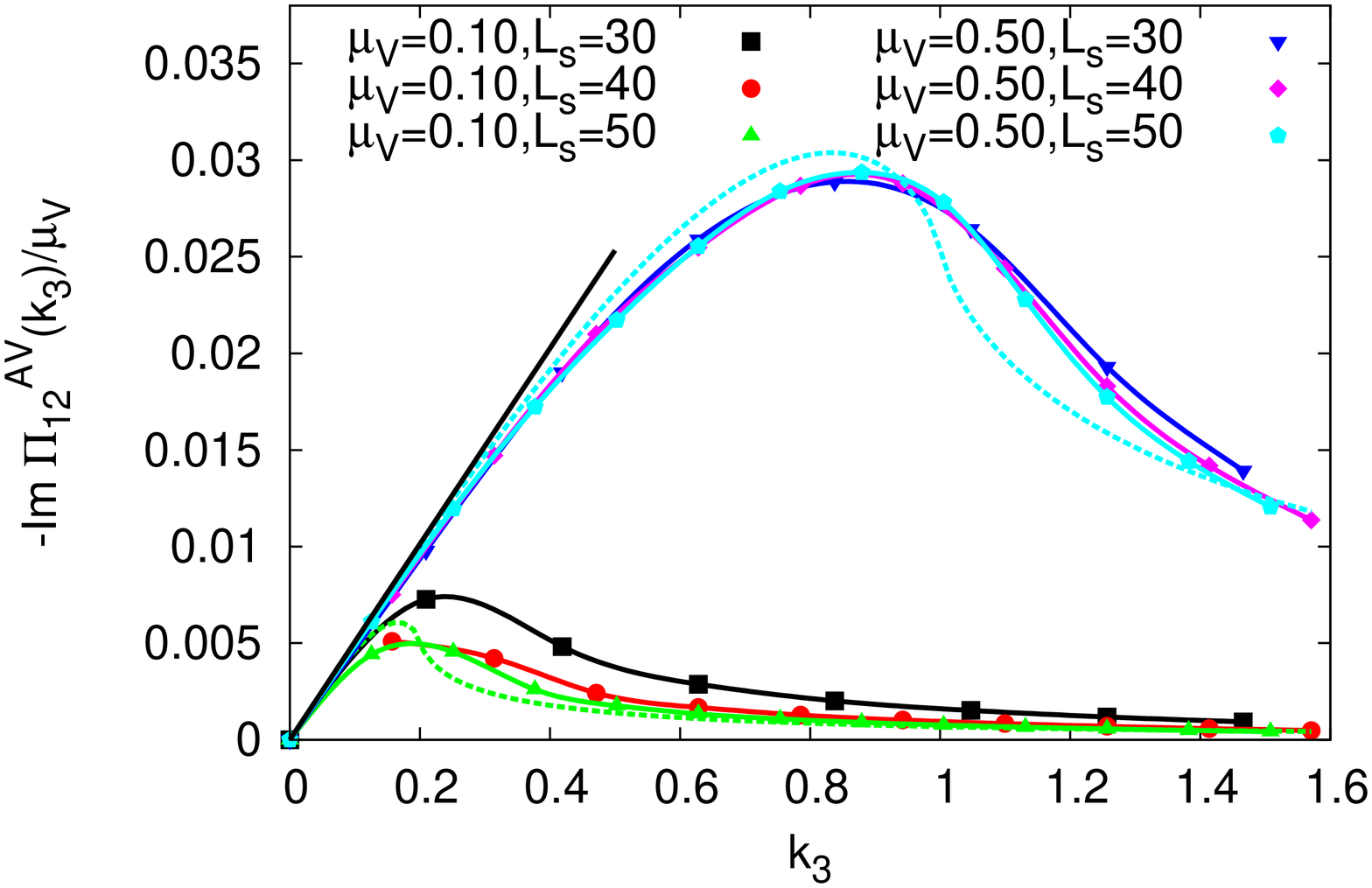}\\
 \caption{Off-diagonal components of the correlator of axial and vector currents $\Pi_{12}^{AV}\lr{k_3}$ for overlap fermions at nonzero chemical potential $\mu_V$. On the left: at different values of the chemical potential $\mu_V$ on $50^4$ lattice, on the right: at $\mu_V = 0.10$ and $\mu_V = 0.50$ and at different lattice sizes. Points connected with solid lines correspond to the lattice result and dashed lines correspond to the continuum expression (\ref{CSEPolTens}).}
 \label{fig:cse_4d_cons}
\end{figure*}

 To study the Chiral Separation Effect, we calculate the correlator of axial and vector currents (\ref{av_correlator_def}) using free overlap lattice fermions at nonzero chemical potential $\mu_V$. Overlap Dirac operator at finite chemical potential is defined as \cite{Bloch:06:1}:
\begin{eqnarray}
\label{overlap_finite_mu}
 \mathcal{D}_{ov}\lr{\mu_V} = 1 + \gamma_5 \sign\lr{\gamma_5 \mathcal{D}_w\lr{\mu_V} }  ,
\end{eqnarray}
where $\mathcal{D}_w$ is the Wilson-Dirac operator at finite chemical potential $\mu_V$ and we have assumed that the lattice spacing is equal to unity. The operator $\mathcal{H} \equiv \gamma_5 \mathcal{D}_w\lr{\mu_V}$ is in general non-Hermitean. The operator-valued sign function of $H$ is then defined in terms of the spectral decomposition of $H$ as
\begin{eqnarray}
\label{sign_nonhermitean}
 \sign\lr{\mathcal{H}} = \sum\limits_i \sign\lr{\re \lambda_i} \, \ket{R_i} \bra{L_i} ,
\end{eqnarray}
where the right and left eigenvectors $\ket{R_i}$ and $\bra{L_i}$ are defined by the equations
\begin{eqnarray}
\label{lr_decomposition_general}
 \mathcal{H} \ket{R_i} = \lambda_i \ket{R_i} , \quad \bra{L_i} \mathcal{H} = \lambda_i \bra{L_i}
 \nonumber \\
 \bra{L_i} \ket{R_j} = \delta_{ij} .
\end{eqnarray}

 Another component which we need in order to calculate the axial-vector correlator (\ref{av_correlator_observable}) is the derivative $\partial^A_{x, \mu}$ over the axial gauge field. As discussed in \cite{Kikukawa:98:1}, for overlap Dirac operator the derivative over $A_{x, \mu}$ is related to the derivative over $V_{x, \mu}$ in a simple way:
\begin{eqnarray}
\label{kikukawa_equation}
 \partial^A_{x, \mu} \mathcal{D}_{ov}
  =
 \partial^V_{x, \mu} \mathcal{D}_{ov} \gamma_5 \lr{1 - \mathcal{D}_{ov}} .
\end{eqnarray}
Technical details of the calculation of the derivatives $\partial^V_{x, \mu} \mathcal{D}_{ov}\lr{\mu}$, $\partial^A_{x, \mu} \mathcal{D}_{ov}\lr{\mu_V}$ and $\partial^V_{y, \nu} \partial^A_{x, \mu} \mathcal{D}_{ov}\lr{\mu_V}$ which enter the expression (\ref{av_correlator_observable}) are not crucial for the understanding of our results and are given in Appendix \ref{apdx:overlap_derivatives}. In short, we have expressed the derivatives of $\mathcal{D}_{ov}\lr{\mu_V}$ in terms of the derivatives of the eigenvectors and eigenvalues of $\mathcal{H}$, which can be in turn related to the derivatives $\partial^V_{x, \mu} \mathcal{H}$, $\partial^A_{x, \mu} \mathcal{H}$ and $\partial^V_{y, \nu} \mathcal{H}$ of the operator $\mathcal{H}$ itself. Since $\mathcal{H}$ is a local lattice operator which is known explicitly, these derivatives can be calculated in a straightforward way. It should be stressed that while for free lattice fermions the spectrum of $\mathcal{H}$ is known analytically, even in this case the exact numerical calculation of the derivatives $\partial^V_{x, \mu} \mathcal{D}_{ov}\lr{\mu_V}$, $\partial^A_{x, \mu} \mathcal{D}_{ov}\lr{\mu_V}$ and $\partial^V_{y, \nu} \partial^A_{x, \mu} \mathcal{D}_{ov}\lr{\mu_V}$ turns out to take a large amount of computer time. For that reason we were only able to compute the current-current correlators on the lattices with sizes not higher than $50^4$. Such computational complexity even for free fermions clearly calls for some efficient way to numerically approximate the conserved vector and axial current vertices for overlap Dirac operator using, e.g., the Krylov subspace methods. We are now studying possible algorithms for sufficiently fast calculation of derivatives, but the discussion of them is out of the scope of the present paper.

 Off-diagonal component of the correlator of axial and vector currents $\Pi_{12}^{AV}\lr{k_3}$ computed with overlap fermions at finite chemical potential $\mu_V$ is plotted on Fig. \ref{fig:cse_4d_cons} as a function of spatial momentum $k_3$. Points with solid lines plotted through them correspond to the lattice result. Solid lines which connect the data points are only intended to guide the eye and are simply cubic splines. Dashed lines correspond to the continuum expression (\ref{CSEPolTens}). The left plot on Fig. \ref{fig:cse_4d_cons} shows the function $\Pi_{12}^{AV}\lr{k_3}$ on the largest lattice size which we have used, $50^4$, and at different values of the chemical potential $\mu_V$. We find a good agreement with the continuum result for not very large values of chemical potential (much smaller than the lattice spacing $a$, which is equal to unity in our case). The right plot on Fig. \ref{fig:cse_4d_cons} illustrates the finite-size effects in $\Pi_{12}^{AV}\lr{k_3}$ by showing the function  $\Pi_{12}^{AV}\lr{k_3}$ at two fixed values of chemical potential, $\mu_V=0.10$ and $\mu_V=0.50$ and at different lattice sizes. One can see that indeed as the lattice size becomes larger, the lattice result approaches the continuum expression (\ref{CSEPolTens}).

\subsection{Chiral Magnetic Effect}
\label{subsec:cme_overlap}

\begin{figure*}[Htpb]
\includegraphics[width=5cm, angle=-90]{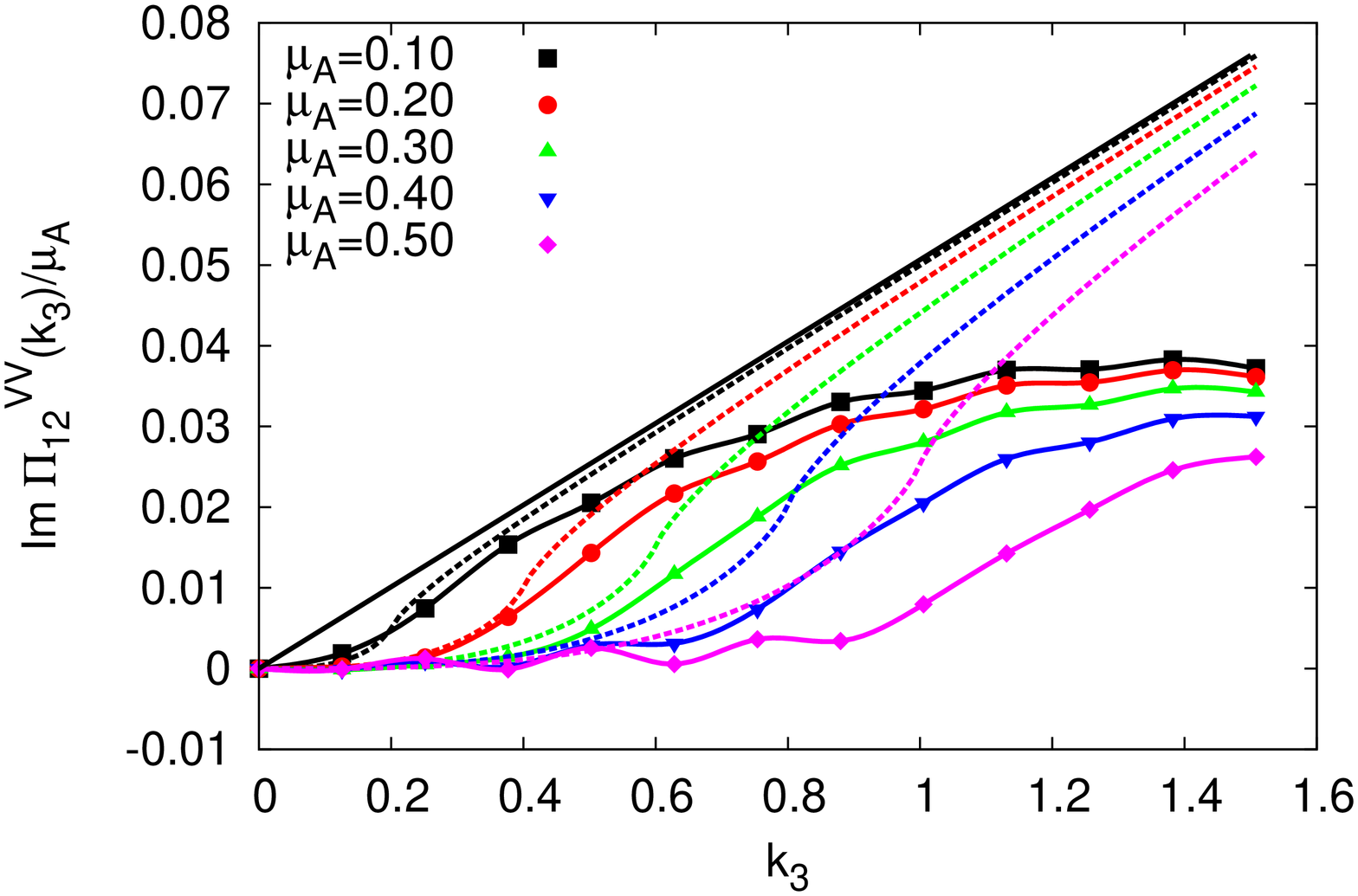}
\includegraphics[width=5cm, angle=-90]{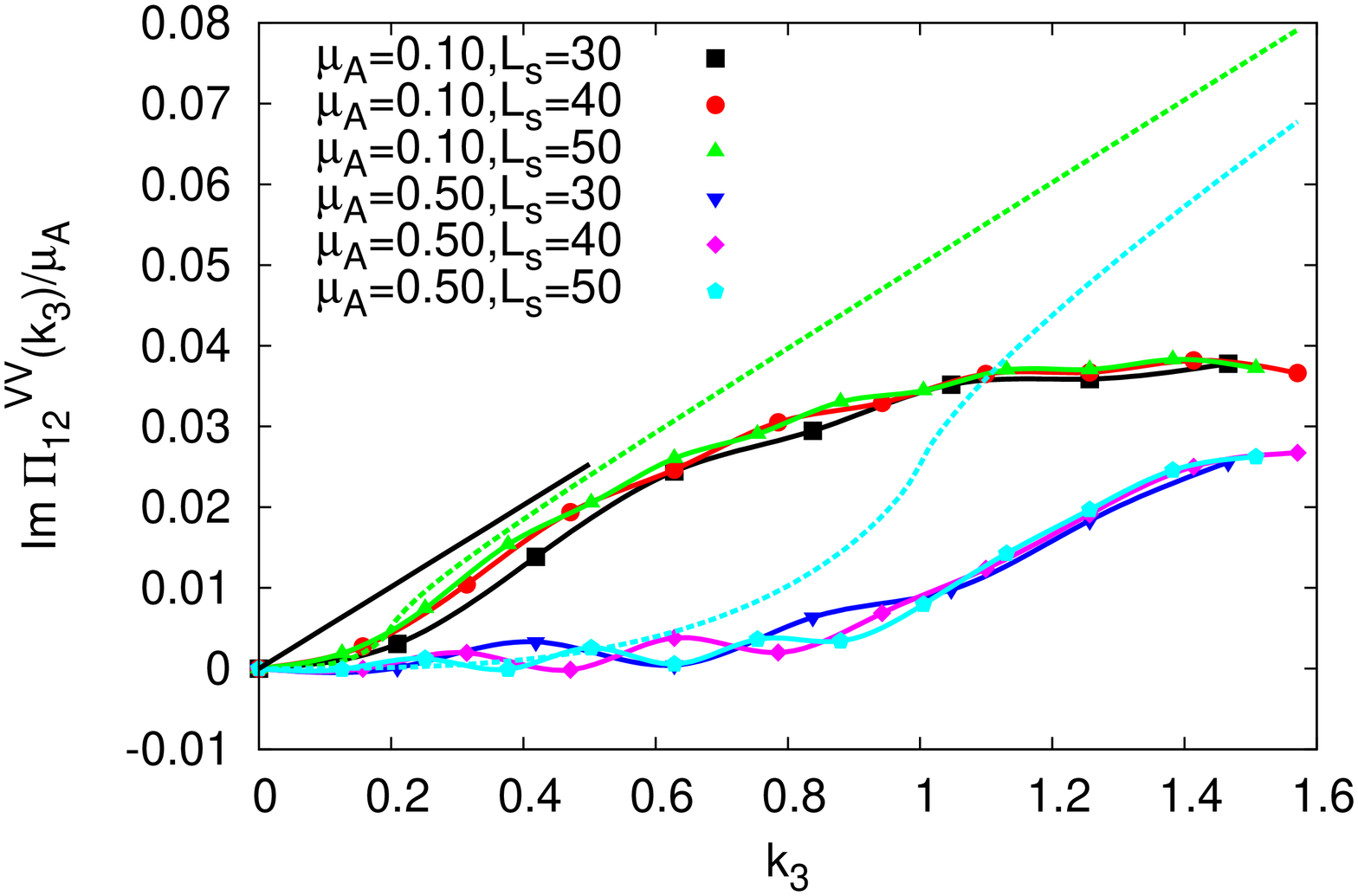}\\
 \caption{Off-diagonal components of the correlator of two vector currents $\Pi_{12}^{VV}\lr{k_3}$ for overlap fermions at nonzero chiral chemical potential $\mu_A$. On the left: at different values of the chemical potential on $50^4$ lattice, on the right: at $\mu_A = 0.10$ and $\mu_A = 0.50$ and at different lattice sizes. Points connected with solid lines correspond to the lattice result and dashed lines correspond to the continuum expression (\ref{CMEPolTens}).}
 \label{fig:cme_4d_cons}
\end{figure*}

 In this Subsection we consider the correlator of two vector currents (\ref{vv_correlator_observable}) at nonzero chiral chemical potential $\mu_A$, which describes the Chiral Magnetic Effect. Overlap Dirac operator with chiral chemical potential has been constructed recently in \cite{Buividovich:13:6}. It can be implicitly expressed in terms of overlap Dirac operator at nonzero chemical potential $\mu_V$ (see equations (\ref{overlap_finite_mu}) and (\ref{sign_nonhermitean})) as
\begin{eqnarray}
\label{overlap_mu5_implicit}
 \tilde{\mathcal{D}}_{ov}\lr{\mu_A}
 = 
 \mathcal{P}_{-} \tilde{\mathcal{D}}_{ov}\lr{\mu_V=+\mu_A} \mathcal{P}_{+}
 + \nonumber \\ +
 \mathcal{P}_{+} \tilde{\mathcal{D}}_{ov}\lr{\mu_V=-\mu_A} \mathcal{P}_{-} ,
\end{eqnarray}
where $\mathcal{P}_{\pm} = \frac{1 \pm \gamma_5}{2}$ are the chiral projectors and
\begin{eqnarray}
\label{gw_projection}
 \tilde{\mathcal{D}}_{ov} = \frac{2 \mathcal{D}_{ov}}{2 - \mathcal{D}_{ov}} = -2 + \frac{4}{2 - \mathcal{D}_{ov}}
\end{eqnarray}
is the ``projected'' overlap Dirac operator. Such projection conformally maps the Ginsparg-Wilson circle in the complex plane of Dirac eigenvalues onto imaginary axis. The inverse projection is
\begin{eqnarray}
\label{inverse_gw_projection}
 \mathcal{D}_{ov} = \frac{2 \tilde{\mathcal{D}}_{ov}}{2 + \tilde{\mathcal{D}}_{ov}} = 2 - \frac{4}{2 + \tilde{\mathcal{D}}_{ov}} .
\end{eqnarray}
In short, one arrives at the representation (\ref{overlap_mu5_implicit}) in the following way \cite{Buividovich:13:6}: first, one observes that the chiral chemical potential couples to fermions as the imaginary time-like component of the axial lattice gauge field $A_{x, \mu}$. Then one requires that gauge transformations of the axial gauge field $A_{x, \mu} \rightarrow A_{x, \mu} + \delta\theta_{x+\hat{\mu}} - \delta\theta_x$ generate the local L\"{u}scher transformations of the overlap Dirac operator $\mathcal{D}_{ov}$:
\begin{eqnarray}
\label{luscher_transform}
 \lrs{\delta \mathcal{D}_{ov}}_{xy}
 =
 \sum\limits_z\lrs{1 - \mathcal{D}_{ov}/2}_{xz} \gamma_5 \delta\theta_z \lrs{\mathcal{D}_{ov}}_{zy}
 + \nonumber \\ +
 \sum\limits_z\lrs{\mathcal{D}_{ov}}_{xz} \delta\theta_z \gamma_5 \lrs{1 - \mathcal{D}_{ov}/2}_{zy} .
\end{eqnarray}

 It is easy to check that the overlap Dirac operator defined by (\ref{overlap_mu5_implicit}) is $\gamma_5$-hermitian and satisfies the Ginsparg-Wilson relations. Thus the eigenvalues of this operator lie on the Ginsparg-Wilson circle in the complex plane, just as for the ordinary overlap operator without any chemical potentials. In particular, these properties imply that the absence of the sign problem in Monte-Carlo simulations with overlap fermions at finite chiral chemical potential.

 In order to calculate the vector-vector correlator (\ref{vv_correlator_observable}) we need the derivatives $\partial^V_{x, \mu} \mathcal{D}_{ov}\lr{\mu_A}$ and $\partial^V_{y, \nu} \partial^V_{x, \mu} \mathcal{D}_{ov}\lr{\mu_A}$. By differentiating the expressions (\ref{overlap_mu5_implicit}), (\ref{gw_projection}) and (\ref{inverse_gw_projection}) we can relate them to the derivatives $\partial^V_{x, \mu} \mathcal{D}_{ov}\lr{\mu_V = \pm \mu_A}$ and $\partial^V_{y, \nu} \partial^V_{x, \mu} \mathcal{D}_{ov}\lr{\mu_V = \pm \mu_A}$. Explicit expressions for $\partial^V_{x, \mu} \mathcal{D}_{ov}\lr{\mu_A}$ and $\partial^V_{y, \nu} \partial^V_{x, \mu} \mathcal{D}_{ov}\lr{\mu_A}$ turn out to be quite lengthy and are summarized in Appendix \ref{apdx:overlap_derivatives}.

 Off-diagonal component of the correlator of two vector currents $\Pi_{12}^{VV}\lr{k_3}$ computed with overlap fermions at finite chiral chemical potential $\mu_A$ is plotted on Fig. \ref{fig:cme_4d_cons} as a function of spatial momentum $k_3$. Points with solid lines plotted through them correspond to the lattice result. Solid lines which connect the data points are only intended to guide the eye and are simply cubic splines. Dashed lines correspond to the continuum expression (\ref{CMEPolTens}). The left plot on Fig. \ref{fig:cme_4d_cons} shows the function $\Pi_{12}^{VV}\lr{k_3}$ on the largest lattice size which we have used, $50^4$, and at different values of the chiral chemical potential $\mu_A$. We find a good agreement with the continuum result (\ref{CMEPolTens}) for $\mu_A$ and $k_3$ much smaller than the lattice spacing (which is equal to unity in our case). The right plot on Fig. \ref{fig:cme_4d_cons} illustrates the finite-size and finite-volume effects in $\Pi_{12}^{VV}\lr{k_3}$ by showing the function  $\Pi_{12}^{VV}\lr{k_3}$ at two fixed values of chiral chemical potential, $\mu_A=0.10$ and $\mu_A=0.50$ and at different lattice sizes. One can see that indeed as the lattice size becomes larger, at small momenta the lattice result approaches the continuum expression (\ref{CMEPolTens}) obtained in the Pauli-Villars regularization. We see once again that it is not correct to simply add the contributions of different chiral states before integrating over the spatial momentum in (\ref{CMEPolTens3}). Rather, the contribution from each chiral sector should be regularized individually, as it is done in the overlap formalism \cite{Neuberger:93:1, Slavnov:93:1}. Upon such a regularization the chiral magnetic conductivity turns out to be zero.

 In agreement with the expression (\ref{CMEPolTens}) derived in the continuum theory, at large momenta the anomalous vector-vector correlator behaves as $\Pi_{12}^{VV}\lr{k_3} = \frac{i k_3 \mu_A}{2 \pi^2}$, which coincides up to a sign with the naive conventional asymptotic behavior (\ref{CMEClassicPolTens}) at small momenta. In Section \ref{sec:rel_to_anomaly} below we will demonstrate that this is not an accidental coincidence. Rather, the slope of $\Pi_{12}^{VV}\lr{k_3}$ turns out to be related to the anomaly coefficient. From Fig.~\ref{fig:cme_4d_cons} we see that in order to reproduce this asymptotically linear behavior, one needs to perform measurements for momenta $k_3 \gg \mu_A$. In turn, both $k_3$ and $\mu_A$ should be much smaller than the inverse lattice spacing. Then quite large lattice size is required in order to have a sufficient number of discrete lattice momenta in the range $\mu_A \ll k_3 \ll a^{-1}$.

\begin{figure*}[Htpb]
\includegraphics[width=5cm, angle=-90]{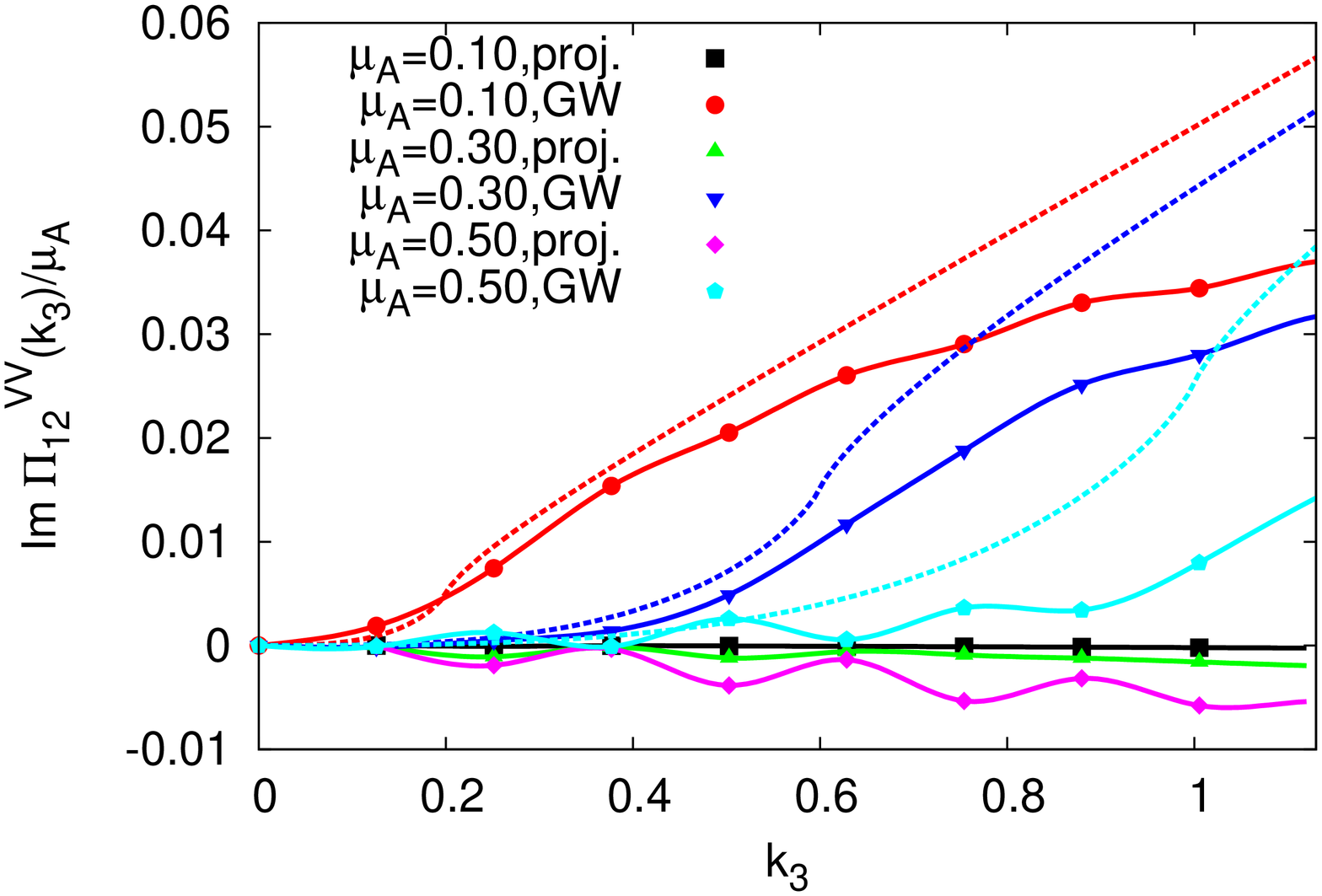}
\includegraphics[width=5cm, angle=-90]{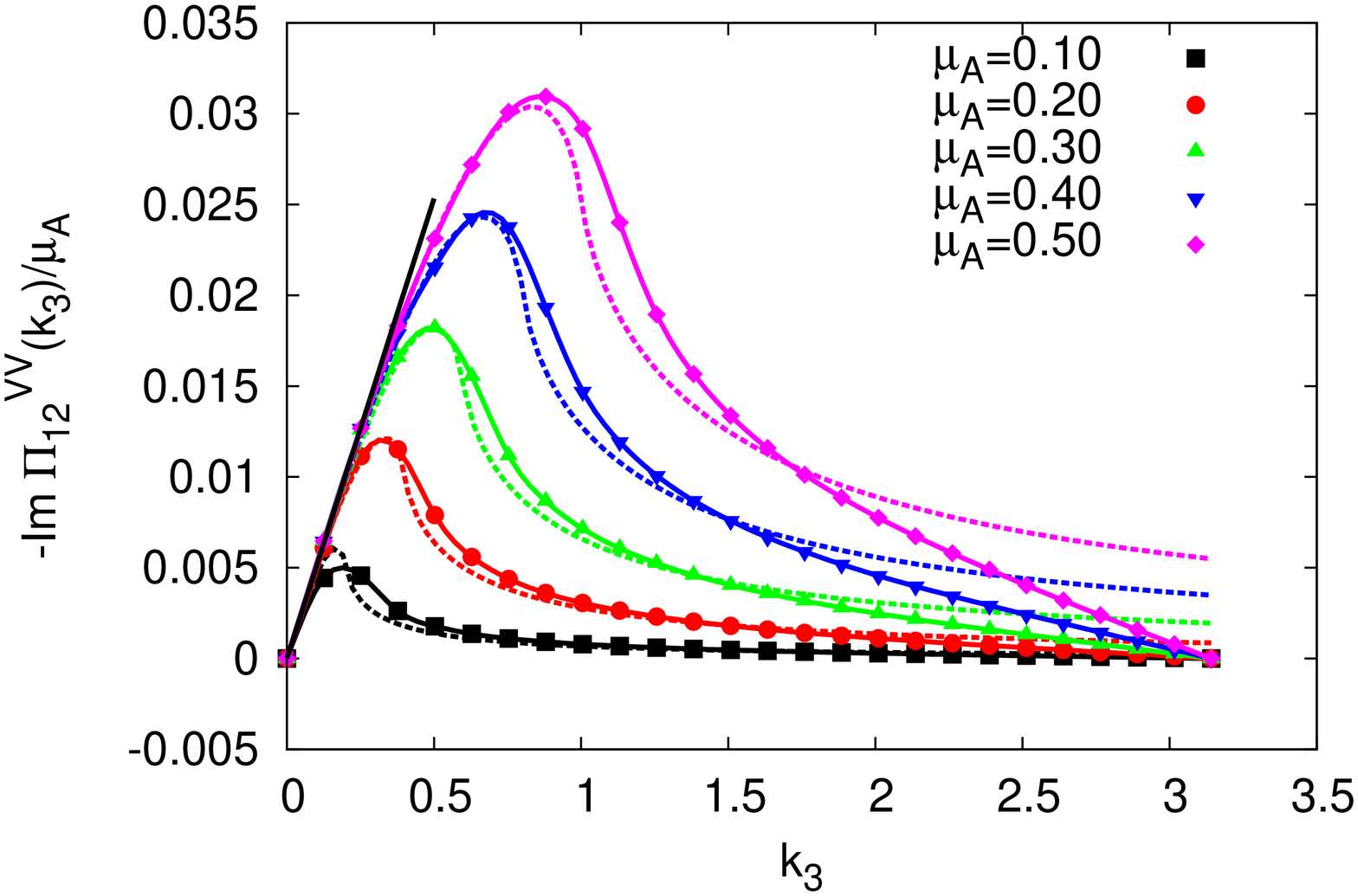}\\
 \caption{Off-diagonal components of the correlator of two vector currents $\Pi_{12}^{VV}\lr{k_3}$ for overlap fermions at nonzero chiral chemical potential $\mu_A$ on the $50^4$ lattice. On the left: using the projected overlap Dirac operator $\tilde{\mathcal{D}}_{ov}\lr{\mu_A}$ (\ref{overlap_mu5_implicit}) instead of $\mathcal{D}_{ov}\lr{\mu_A}$, compared with the results obtained with $\mathcal{D}_{ov}\lr{\mu_A}$ (label ``GW'') and with the continuum result (\ref{CMEPolTens}) (shown with dashed lines). On the right: the same with non-conserved vector current vertices. Points connected with solid lines correspond to the lattice result and dashed lines correspond to the continuum expression (\ref{CSEPolTens}) with the replacement $\mu_V \rightarrow \mu_A$.}
 \label{fig:cme_4d_pathological}
\end{figure*}

 It is also interesting to check how robust is the result (\ref{CMEPolTens}) against lattice artifacts. First we check how important is it to use the L\"{u}scher transformations (\ref{luscher_transform}) instead of the conventional chiral rotations $\delta \psi_x = \theta_x \gamma_5 \psi_x$, $\delta \bar{\psi}_x = \theta_x \bar{\psi_x} \gamma_5$. To this end we simply insert the projected overlap Dirac operator $\tilde{\mathcal{D}}_{ov}\lr{\mu_A}$ (\ref{overlap_mu5_implicit}) instead of $\mathcal{D}_{ov}\lr{\mu_A}$ into the definition (\ref{av_correlator_observable}) of the axial-vector correlator (\ref{av_correlator_def}) on the lattice. It is easy to check that local L\"{u}scher transformations (\ref{luscher_transform}) of the operator $\mathcal{D}_{ov}$ result in the conventional local chiral rotations of the projected operator $\tilde{\mathcal{D}}_{ov}$:
\begin{eqnarray}
\label{local_chiral_rotations}
 \lrs{\delta \tilde{\mathcal{D}}_{ov}}_{xy} = \delta\theta_x \gamma_5 \lrs{\tilde{\mathcal{D}}_{ov}}_{xy} + \lrs{\tilde{\mathcal{D}}_{ov}}_{xy} \delta\theta_y \gamma_5  .
\end{eqnarray}
At small momenta, both operators $\mathcal{D}_{ov}$ and $\tilde{\mathcal{D}}_{ov}$ reproduce the conventional continuum lattice Dirac operator with chiral chemical potential. Thus in the continuum limit both operators should be equivalent, and any difference in the observables involving them comes from the effects of ultraviolet cutoff, that is, from the details of the lattice regularization. The vector-vector correlator $\Pi_{12}^{VV}\lr{k_3}$ which was calculated using the projected operator $\tilde{\mathcal{D}}_{ov}$ is plotted on Fig. \ref{fig:cme_4d_pathological} on the left. For comparison, we also plot the results obtained with the operator $\mathcal{D}_{ov}$, which correctly incorporates lattice chiral symmetry. The result which we get with $\tilde{\mathcal{D}}_{ov}$ does not resemble the regularized continuum result (\ref{CSEPolTens}) at all and is almost by an order of magnitude smaller. We can explain this discrepancy as follows: in the case when local chiral transformations are given by simple rotations (\ref{local_chiral_rotations}), the axial anomaly comes from the nontrivial regularization of the jacobian of the transformations (\ref{local_chiral_rotations}). Obviously, in a finite volume and on a finite lattice the jacobian of (\ref{local_chiral_rotations}) is always trivial (see e.g. \cite{Nielsen:83:1, Teryaev:93:1}). For lattice Dirac operators which satisfy the Ginsparg-Wilson relation the anomaly is reproduced even on finite-volume lattices due to the nontrivial jacobian of the L\"{u}scher transformations (\ref{luscher_transform}). We conclude that in order to correctly reproduce the Chiral Magnetic Effect it is essential to have a nontrivial jacobian of chiral rotations, which is directly related to the axial anomaly.

 Another commonly used approximation for the calculation of current-current correlators with overlap fermions is the use of the Dirac gamma-matrices for the definition of the current operator $j_{x, \mu} = \bar{\psi}_x \gamma_{\mu} \psi_x$. In other words, one approximates the vector-vector correlator (\ref{vv_correlator_observable}) as
\begin{eqnarray}
\label{vv_correlator_noncons}
 \vev{j^V_{x, \mu} j^V_{y, \nu}}
 \approx
 T^2 \, N_f \, \vev{ \tr\lr{\mathcal{D}^{-1}_{ov\,yx} \gamma_{\mu} \mathcal{D}^{-1}_{ov\,xy} \gamma_{\nu} } }
 - \nonumber \\ -
 T^2 \, N_f^2 \, \vev{ \tr\lr{\mathcal{D}^{-1}_{ov\,xx} \gamma_{\mu} } \tr\lr{\mathcal{D}^{-1}_{ov\,yy} \gamma_{\nu}} }   .
\end{eqnarray}
The current defined in this way is only conserved up to lattice artifacts which by naive power counting should vanish in the continuum limit: $\partial_{\mu} j^V_{x, \mu} = O\lr{a^2}$, where $a$ is the lattice spacing. Correspondingly, the correlator (\ref{vv_correlator_noncons}) satisfies the vector Ward identities only approximately. In the above expressions, one can also use the projected Dirac operator defined by (\ref{inverse_gw_projection}). In order to see how such current non-conservation affects the chiral magnetic conductivity, we have calculated the vector-vector correlator (\ref{vv_correlator_noncons}) using the projected overlap Dirac operator at finite chiral chemical potential defined by (\ref{overlap_mu5_implicit}). The resulting vector-vector correlator $\Pi_{12}^{VV}\lr{k_3}$ is plotted on Fig.~\ref{fig:cme_4d_pathological} on the right for different values of the chiral chemical potential $\mu_A$. Quite surprisingly, using the non-conserved current we almost exactly reproduce the naive unregularized continuum expression (\ref{CMEPolTensMassless}) (it is also plotted on Fig.~\ref{fig:cme_4d_pathological} on the right with dashed lines). This observation clarifies why it is not correct to add up the contributions of different chiral states before integrating over the spatial momentum in (\ref{CMEPolTens3}): such a prescription leads to the current-current correlator which do not satisfy the vector Ward identities. Similar observation was also made recently in \cite{Buividovich:13:6}, where it was found that the Chiral Magnetic Effect (\ref{CMEClassicExpression}) for a constant magnetic field in a finite volume is only realized for non-conserved electric current.

\begin{figure}[Htpb]
 \includegraphics[width=5cm, angle=-90]{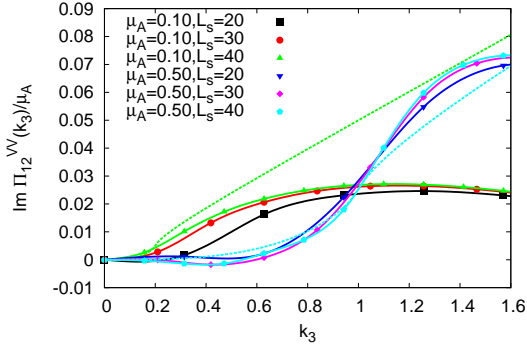}\\
 \caption{Off-diagonal components of the correlator of two vector currents $\Pi_{12}^{VV}\lr{k_3}$ for massless Wilson-Dirac fermions at nonzero chiral chemical potential $\mu_A$ for different values of $\mu_A$ and different lattice sizes. Points connected with solid lines correspond to the lattice result and dashed lines correspond to the continuum expression (\ref{CMEPolTens}).}
 \label{fig:cme_4d_wdirac}
\end{figure}

 Finally, it is interesting to see how the Chiral Magnetic Effect is reproduced for lattice fermions without exact chirality. To this end we calculate the current-current correlator (\ref{vv_correlator_observable}) for free massless Wilson-Dirac fermions. In order to couple the Wilson-Dirac operator to the chiral chemical potential, we simply replace the factors $e^{\pm \mu_V}$ on the time-like links in the Wilson-Dirac operator at finite chemical potential $\mu_V$ (see the expression (\ref{WDOperator}) in Appendix \ref{apdx:overlap_derivatives}) with matrix exponents $e^{\pm \gamma_5 \mu_A} = \cosh\lr{\mu_A} \pm \gamma_5 \sinh\lr{\mu_A}$. The resulting current-current correlator is shown on Fig.~\ref{fig:cme_4d_wdirac} as a function of the spatial momentum $k_3$ for different values of the chiral chemical potential $\mu_A$ and for different lattice sizes. Dashed lines on the plot correspond to the continuum expression (\ref{CMEPolTens}). One can see that for $\mu_A = 0.10$, which is significantly smaller than the lattice spacing $a = 1$, the lattice result approaches the continuum result as the lattice size increases. For larger value of $\mu_A$, $\mu_A = 0.50$, the agreement is not so good, thus the finite-spacing artifacts are quite large. We conclude that the expression (\ref{CMEPolTens}) is reproduced even for Wilson-Dirac fermions, although with much larger discretization artifacts than for overlap fermions.

 From the results presented in this Section, we conclude that a regularization of individual contributions of states with different chiralities is necessary in order to get a consistent result for the chiral magnetic conductivity. In any regularization with conserved vector current it turns out to be equal to zero, in agreement with the arguments of \cite{Rubakov:10:1, Ren:11:1}.

\section{Relation to axial anomaly}
\label{sec:rel_to_anomaly}

 In this Section we analyze more carefully the relation of the Chiral Separation and Chiral Magnetic Effects to the axial anomaly. In contrast to the previous Sections, where we have considered free fermions, the results presented in this Section are also valid in interacting theories.

 At first sight the relation of the anomalous current-current correlators (\ref{CSEPolarizationTensor}) and (\ref{CMEPolarizationTensor}) to axial anomaly is far from obvious, since the conventional anomaly relation involves the divergence $\partial_{\mu} j^A_{\mu}\lr{x}$ rather than the axial-vector and the vector-vector correlators (\ref{CSEPolarizationTensor}) and (\ref{CMEPolarizationTensor}). Let us, however, formally expand the vector-vector and the axial-vector correlators in powers of the corresponding chemical potentials and consider the linear terms in such expansions, which are given by the derivatives
\begin{eqnarray}
 \label{av_correlator_mu_exp}
 \frac{\partial}{\partial \mu_V} \Pi_{12}^{AV}\lr{k_3}|_{\mu_V = 0} \equiv -\Gamma_{201}^{VVA}\lr{0, k_3}
 = \nonumber \\ =
 -\int d^4y \int d^4x \, e^{i k_3 x_3} \, \vev{ j^A_1\lr{x} j^V_2\lr{0} j^V_0\lr{y}} ,
 \\
 \label{vv_correlator_mu_exp}
 \frac{\partial}{\partial \mu_A} \Pi_{12}^{VV}\lr{k_3}|_{\mu_A = 0} \equiv -\Gamma_{120}^{VVA}\lr{k_3, -k_3}
 = \nonumber \\ =
 -\int d^4y \int d^4x \, e^{i k_3 x_3} \, \vev{ j^V_1\lr{x} j^V_2\lr{0} j^A_0\lr{y}} .
\end{eqnarray}
In deriving the above expressions, we have used the fact that the first derivative of the current-current correlators (\ref{CSEPolarizationTensor}) and (\ref{CMEPolarizationTensor}) over the corresponding chemical potentials simply brings down minus the total electric or axial charge $-T^{-1} \int d^3\vec{x} j^V_0\lr{\vec{x}} = -\int d^4x j^V_0\lr{x}$ and $-T^{-1} \int d^3\vec{x} j^A_0\lr{\vec{x}} = -\int d^4x j^A_0\lr{x}$ into the expectation values in (\ref{CSEPolarizationTensor}) and (\ref{CMEPolarizationTensor}). For notational convenience, we have also defined a generic correlator of two vector and one axial currents $\Gamma_{\mu\nu\rho}^{VVA}\lr{p_{\mu}, q_{\mu}}$:
\begin{eqnarray}
\label{vva_correlator_def}
 \Gamma_{\mu\nu\rho}^{VVA}\lr{p, q}
 =
 \int d^4x \int d^4y \,
 e^{i p x + i q y}
 \times \nonumber \\ \times
 \vev{j^V_{\mu}\lr{x} j^V_{\nu}\lr{y} j^A_{\rho}\lr{0}}  .
\end{eqnarray}
This correlator can be directly related to the usual triangle diagram which yields the anomaly coefficient. To be more precise, it can be expressed in terms of four independent tensor structures $t^L_{\mu\nu\rho}\lr{p, q}$, $t_{\mu\nu\rho}^{\lr{\pm}}\lr{p, q}$ and $\tilde{t}_{\mu\nu\rho}\lr{p, q}$ which are allowed by the vector and the anomalous Ward identities and the four corresponding form-factors $w_L\lr{p^2, q^2, \lr{p+q}^2}$, $w_T^{\lr{\pm}}\lr{p^2, q^2, \lr{p + q}^2}$ and $\tilde{w}_T\lr{p^2, q^2, \lr{p + q}^2}$ \cite{Knecht:04:1}:
\begin{widetext}
\begin{eqnarray}
\label{VVA_general_decomposition}
 \Gamma^{VVA}_{\mu\nu\rho}\lr{p, q} = \frac{i}{4 \pi^2} \left(
 w_L\lr{p^2, q^2, \lr{p + q}^2} t^L_{\mu\nu\rho}\lr{p, q}
 +
 \tilde{w}_T\lr{p^2, q^2, \lr{p + q}^2} \tilde{t}_{\mu\nu\rho}\lr{p, q}
 + \right. \nonumber \\ \left. +
 w_T^{\lr{+}}\lr{p^2, q^2, \lr{p + q}^2} t^{\lr{+}}_{\mu\nu\rho}\lr{p, q}
 +
 w_T^{\lr{-}}\lr{p^2, q^2, \lr{p + q}^2} t^{\lr{-}}_{\mu\nu\rho}\lr{p, q}
 \right)  ,
\end{eqnarray}
where
\begin{eqnarray}
\label{VVA_tensors}
 t^L_{\mu\nu\rho}\lr{p, q} = -\lr{p + q}_{\rho} \epsilon_{\mu \nu \alpha \beta} p_{\alpha} q_{\beta}
 \nonumber \\
 t^{\lr{+}}_{\mu\nu\rho}\lr{p, q}
 =
 p_{\nu} \epsilon_{\mu \rho \alpha \beta} p_{\alpha} q_{\beta}
 -
 q_{\mu} \epsilon_{\nu \rho \alpha \beta} p_{\alpha} q_{\beta}
 -
 \lr{p \cdot q} \epsilon_{\mu \nu \rho \alpha} \lr{p - q}_{\alpha}
 - \nonumber \\ -
 \frac{2 p \cdot q}{\lr{p + q}^2} \epsilon_{\mu\nu\alpha\beta} p_{\alpha} q_{\beta} \lr{p + q}_{\rho}
 \nonumber \\
 t^{\lr{-}}_{\mu\nu\rho}\lr{p, q}
 =
 \lr{\lr{p - q}_{\rho} - \frac{p^2 - q^2}{\lr{p + q}^2} \lr{p + q}_{\rho} }
 \epsilon_{\mu\nu\alpha\beta} p_{\alpha} q_{\beta}
 \nonumber \\
 \tilde{t}_{\mu\nu\rho}\lr{p, q}
 =
 p_{\nu} \epsilon_{\mu\rho\alpha\beta} p_{\alpha} q_{\beta}
 +
 q_{\mu} \epsilon_{\nu\rho\alpha\beta} p_{\alpha} q_{\beta}
 -
 \lr{p \cdot q} \epsilon_{\mu \nu \rho \alpha} \lr{p + q}_{\alpha}  .
\end{eqnarray}
\end{widetext}
Here we have adapted the original expressions given in \cite{Knecht:04:1} to the case of flavour-singlet currents in Euclidean space and assumed that the number of quark colours is equal to one (as in the rest of the paper). The form-factor $w_L\lr{q_1^2, q_2^2, \lr{q_1 + q_2}^2}$ is directly related to the axial anomaly coefficient and does not receive neither perturbative nor non-perturbative corrections:
\begin{eqnarray}
\label{AnomalyFormFactor}
 w_L\lr{p^2, q^2, \lr{p + q}^2} = -\frac{2}{\lr{p + q}^2} .
\end{eqnarray}
It should be stressed that the expression (\ref{VVA_general_decomposition}) together with (\ref{AnomalyFormFactor}) already take into account all information which is contained in both vector and axial Ward identities, and the transverse form-factors $w_T^{\lr{\pm}}$ and $\tilde{w}_T$ cannot be fixed from anomaly equations.

 Before proceeding with the discussion of the relation of the derivatives (\ref{av_correlator_mu_exp}) and (\ref{vv_correlator_mu_exp}) to the axial anomaly, let us clarify their relation to the asymptotic behavior of the functions $\Pi_{12}^{AV}\lr{k_3}$ and $\Pi_{12}^{VV}\lr{k_3}$ in the case of free massless continuum fermions. To this end we represent the derivatives (\ref{av_correlator_mu_exp}) and (\ref{vv_correlator_mu_exp}) as the following limit:
\begin{eqnarray}
\label{derivative_def}
 \frac{\partial}{\partial \mu} \Pi_{12}\lr{k_3}|_{\mu = 0} = \lim\limits_{\mu \rightarrow 0} \frac{\Pi_{12}\lr{k_3, \mu} - \Pi_{12}\lr{k_3, 0}}{\mu}  ,
\end{eqnarray}
where for generality we have denoted either the vector-vector or axial-vector correlators as $\Pi_{12}\lr{k_3}$ and the corresponding chemical potential - simply as $\mu$. For free massless fermions the only scale entering the current-current correlation functions is the value of the corresponding chemical potential. Therefore when we take the limit $\mu \rightarrow 0$ in (\ref{derivative_def}), we have to use the asymptotic forms of the functions $\Pi_{12}^{AV}\lr{k_3}$ or $\Pi_{12}^{VV}\lr{k_3}$ in the limit $k_3 \gg \mu_V$ or $k_3 \gg \mu_A$. We thus conclude that in the absence of any intrinsic scale other than the values of the chemical potential, the derivative (\ref{derivative_def}) at fixed momentum $k_3$ is related to the asymptotic behavior of correlation functions in the limit of very large momentum, which is actually irrelevant for hydrodynamics. The origin of such non-commutativity of the limits of small $k_3$ and small $\mu_V$ or $\mu_A$ obviously lies in the non-analyticity of the functions $\Pi_{12}^{AV}\lr{k_3}$ or $\Pi_{12}^{VV}\lr{k_3}$ (given by (\ref{CSEPolTens}) and (\ref{CMEPolTens})) with respect to $k_3$ and $\mu_V$ or $\mu_A$.

 In the case of the Chiral Separation Effect, the asymptotic behavior of $\Pi_{12}^{AV}\lr{k_3}$ at $k_3 \gg \mu_V$ is given by (\ref{CSEPolTensLargeK}). Inserting this expression into (\ref{derivative_def}), we immediately conclude that
\begin{eqnarray}
\label{av_correlator_1st_deriv}
 \frac{\partial}{\partial \mu_V} \Pi_{12}^{AV}\lr{k_3}|_{\mu_V = 0} = 0 .
\end{eqnarray}
For the vector-vector correlator (\ref{CMEPolarizationTensor}) which describes the Chiral Magnetic Effect, we use the asymptotic form $\Pi_{12}^{VV}\lr{k_3} \rightarrow \frac{i k_3 \mu_A}{2 \pi^2}$ of the regularized expression (\ref{CMEPolTens}). It follows immediately that
\begin{eqnarray}
\label{vv_correlator_1st_deriv}
 \frac{\partial}{\partial \mu_A} \Pi_{12}^{VV}\lr{k_3}|_{\mu_A = 0} = \frac{i k_3}{2 \pi^2} .
\end{eqnarray}

 Let us now discuss the relation of the derivatives (\ref{av_correlator_mu_exp}) and (\ref{vv_correlator_mu_exp}) to axial anomaly. First, we take the limit $p = \lr{0, 0, 0, k_3}$, $q = 0$ in the decomposition (\ref{VVA_general_decomposition}) with $\mu = 2$, $\nu = 0$, $\rho = 1$, which corresponds to the linear order (\ref{av_correlator_mu_exp}) of the expansion of the axial-vector correlator in powers of chemical potential $\mu_V$. In full agreement with (\ref{av_correlator_1st_deriv}), the result is zero. We see that the vanishing of the anomalous axial-vector correlator at large spatial momenta is required by vector and axial Ward identities and thus should also hold in interacting theories. In a simple holographic model of \cite{Gynther:10:1}, the vector-vector-axial correlator in the kinematical limit (\ref{av_correlator_mu_exp}) was also found to be zero.

 For the leading order of the expansion of the vector-vector correlator (\ref{vv_correlator_mu_exp}) in powers of chiral chemical potential $\mu_A$, we have to consider the vector-vector-axial correlator (\ref{vva_correlator_def}) in the limit $p = \lr{0, 0, 0, k_3}$, $q = \lr{0, 0, 0, -k_3}$ with $\mu = 1$, $\nu = 2$ and $\rho = 0$. However, a careful analysis of the expressions (\ref{VVA_general_decomposition}), (\ref{VVA_tensors}) and (\ref{AnomalyFormFactor}) shows that the correlator (\ref{vva_correlator_def}) is singular in this limit, and we have to regularize this singularity by allowing for some small momentum $\epsilon = p + q$ flowing in through the axial vertex. To this end we use the parametrization
\begin{eqnarray}
\label{pq_parameterization}
 p = k + \epsilon/2, \quad q = -k + \epsilon/2, \quad k = \lr{0, 0, 0, k_3}.
\end{eqnarray}

 Let us first consider the case when $\epsilon$ has only a temporal component $\epsilon_0$: $\epsilon = \lr{\epsilon_0, 0, 0, 0}$. In this limit the tensor structures $t^{\lr{\pm}}_{\mu\nu\rho}\lr{p, q}$ and $\tilde{t}_{\mu\nu\rho}$ are identically zero, and the only tensor structure in (\ref{VVA_general_decomposition}) which contributes to the derivative (\ref{vv_correlator_mu_exp}) is
\begin{eqnarray}
\label{VVA_tensors_cme_time_dependent}
 t^L_{120} = k_3 \epsilon_0^2 .
\end{eqnarray}
We conclude that in such a limit indeed only the form-factor
\begin{eqnarray}
\label{w_L_cme}
 w_L\lr{\lr{k + \epsilon/2}^2, \lr{k - \epsilon/2}^2, \epsilon^2} = -\frac{2}{\epsilon_0^2}
\end{eqnarray}
which is directly related to axial anomaly contributes to the derivative (\ref{vv_correlator_mu_exp}), and from (\ref{VVA_general_decomposition}) we obtain the result which completely agrees with the expression (\ref{vv_correlator_1st_deriv}) obtained for free fermions. The same result for the vector-vector-axial correlator which enters the expansion (\ref{vv_correlator_1st_deriv}) was also obtained in a holographic calculation of \cite{Gynther:10:1}. We conclude that if one takes the limit of static chiral chemical potential by assuming that it very slowly varies in time, the behavior of the vector-vector correlator (\ref{CMEPolarizationTensor}) is indeed related to axial anomaly, although only at large, rather than at small spatial momentum.

 Such a limit seems somewhat artificial from the point of view of lattice simulations. The only sensible physical interpretation of nonzero $\epsilon_0$ is the analytic continuation from the situation in which the chemical potential slowly varies in real Minkowski time. However, such time-dependent chemical potential is a non-stationary perturbation which brings the system out of thermal equilibrium. A direct connection with thermal expectation values which are commonly calculated in lattice simulations is then lost, and one has to perform a painful reconstruction of real-time spectral functions (see e.g. \cite{Meyer:11:1, Asakawa:01:1}). Thus it seems advantageous to stick to the concept of anomalous transport as a phenomenon which happens in an equilibrium state of a system and assume that the chiral chemical potential $\mu_A$ slowly varies in space but is constant in time. Correspondingly, in (\ref{pq_parameterization}) we assume that $\epsilon$ is an infinitely small spatial vector (so that $\epsilon_0 = 0$). In this limit the only nonzero tensor structures in (\ref{VVA_tensors}) are
\begin{eqnarray}
\label{VVA_tensors_cme_static}
 t^{\lr{+}}_{120}\lr{k+\frac{\epsilon}{2}, k-\frac{\epsilon}{2}} = 2 k_3^3 ,
   \nonumber \\
 \tilde{t}_{120}\lr{k+\frac{\epsilon}{2}, k-\frac{\epsilon}{2}} = k_3^2 \epsilon_3  ,
\end{eqnarray}
where we have kept only the leading order terms in $\epsilon$. Inserting these expressions into the general decomposition (\ref{VVA_general_decomposition}), we see that in this limit the derivative (\ref{vv_correlator_1st_deriv}) of the anomalous vector-vector correlator turns out to be related to the transverse form-factors $w_T^{\lr{+}}\lr{\lr{k+\frac{\epsilon}{2}}^2, \lr{k-\frac{\epsilon}{2}}^2, \epsilon^2}$ and $\tilde{w}_T\lr{\lr{k+\frac{\epsilon}{2}}^2, \lr{k-\frac{\epsilon}{2}}^2, \epsilon^2}$ in the limit $\epsilon \rightarrow 0$. From anti-symmetry properties of the form-factor $\tilde{w}_T\lr{p^2, q^2, \lr{p+q}^2} = - \tilde{w}_T\lr{q^2, p^2, \lr{p+q}^2}$ \cite{Knecht:04:1} we conclude that
\begin{eqnarray}
\label{w_tilde_cme}
 \lim\limits_{\epsilon \rightarrow 0} \tilde{w}_T\lr{\lr{k+\frac{\epsilon}{2}}^2, \lr{k-\frac{\epsilon}{2}}^2, \epsilon^2} = 0  ,
\end{eqnarray}
thus the only nontrivial form-factor in (\ref{VVA_general_decomposition}) which contributes to the expansion (\ref{vv_correlator_mu_exp}) is $w_T^{\lr{+}}\lr{k^2, k^2, 0}$. Here we have also taken the limit $\epsilon \rightarrow 0$, assuming that it is nonsingular. Since the decomposition (\ref{VVA_general_decomposition}) already takes into account both vector and anomalous axial Ward identities, the transverse form-factor $w_T^{\lr{+}}\lr{k^2, k^2, 0}$ and hence also the asymptotic behavior of the vector-vector correlator (\ref{CMEPolarizationTensor}) in an interacting theory are not a priori related to axial anomaly.

  However, in the work \cite{Knecht:04:1} (following the preceding works by Vainshtein et al. \cite{Vainshtein:03:1, Vainshtein:03:2}) several new perturbative non-renormalization theorems for the transverse form-factors $w_T^{\lr{\pm}}\lr{q_1^2, q_2^2, \lr{q_1 + q_2}^2}$ and $\tilde{w}_T^{\lr{\pm}}\lr{q_1^2, q_2^2, \lr{q_1 + q_2}^2}$ in massless QCD were proven. In Appendix \ref{apdx:trans_formfact_nonrenorm} we demonstrate that these non-renormalization theorems allow to fix the value
\begin{eqnarray}
\label{w_plus_cme}
 w_T^{\lr{+}}\lr{k^2, k^2, 0} = -\frac{1}{k^2} .
\end{eqnarray}
Inserting this result and also (\ref{w_tilde_cme}) into (\ref{VVA_general_decomposition}), we again confirm the validity of the expression (\ref{vv_correlator_1st_deriv}). It should be stressed that the status of this result is now, however, quite different from the case of chemical potential which slowly varies in time. The validity of the expression (\ref{w_plus_cme}) is only guaranteed in perturbation theory \cite{Knecht:04:1}. Correspondingly, the relation (\ref{vv_correlator_1st_deriv}) now holds only perturbatively and is not protected from non-perturbative corrections. In \cite{Knecht:04:1} it was argued that such corrections might only appear in the chirally broken phase. For this reason it would be particularly interesting to understand how the anomalous current-current correlators (\ref{CMEPolarizationTensor}) and (\ref{CSEPolarizationTensor}) change across the deconfinement phase transition.

 We conclude that in the absence of any scale in the theory except for the values of the chemical potential, the asymptotic behavior of the chiral magnetic and chiral separation conductivities at large spatial momenta can be uniquely fixed from the vector and axial Ward identities at least in perturbation theory. In the case of the axial-vector correlator (\ref{CSEPolarizationTensor}), the behavior at small momenta can also be related to the anomaly by first expanding it in powers of external momentum and then - in powers of chemical potential $\mu_V$ \cite{Son:06:2}. In this case one has to consider the vector-vector-axial correlator (\ref{vva_correlator_def}) at nonzero chemical potential. Since the anomaly is not affected by chemical potential \cite{Gavai:10:1}, the behavior of the anomalous axial-vector correlator (\ref{CSEPolarizationTensor}) at small momenta and hence the chiral separation conductivity turns out to be also directly related to the anomaly coefficient.

\section{Conclusions and discussion}
\label{sec:conclusions}

 In this paper we have studied the anomalous vector-vector and axial-vector current-current correlators which describe the Chiral Magnetic and the Chiral Separation Effects using the truly chiral regularization provided by overlap fermions on the lattice. As a first step towards the study of anomalous transport in interacting theories, we have performed all the calculations for the case of free fermions.

 We have reproduced the conventional result (\ref{CSEClassicPolTens}) for the chiral separation conductivity and found that it is completely free from ultraviolet divergences and regularization ambiguities.

 In contrast, the chiral magnetic conductivity turns out to be zero with a gauge-invariant regularization in which the vector current is conserved, in agreement with the arguments of \cite{Rubakov:10:1, Ren:11:1, Rebhan:10:1}. The conventional result (\ref{CMEClassicPolTens}) can be reproduced if one uses the ``covariant'' vector current instead of the ``consistent'' one which we obtain by differentiating the gauge-invariant partition function (\ref{partition_function_def}) over the background gauge fields \cite{Rebhan:10:1}. Consistent electric current is exactly conserved and is invariant with respect to the gauge transformations of the vector gauge field, but is not invariant under gauge transformations of the axial gauge fields. On the other hand, covariant current is invariant under gauge transformations of both the vector and the axial gauge fields, but it cannot be represented as a derivative of the partition function over the vector gauge field and is therefore not conserved \cite{Rebhan:10:1, Landsteiner:12:1, Landsteiner:13:2}. The difference between the two definitions of the current is again the Chern-Simons term of the form
\begin{eqnarray}
\label{cs_current}
 K_{\mu} = \frac{1}{4 \pi^2} \epsilon_{\mu \nu \rho \sigma} V_{\nu} \partial_{\rho} V_{\sigma}
\end{eqnarray}
with the coefficient $1/\lr{4 \pi^2}$ being completely fixed by the anomaly equation. For different physical situations either the covariant or the consistent definitions of the electric current might be appropriate \cite{Landsteiner:13:2}. For instance, if one uses the covariant electric current, the rate at which the (now non-conserved) electric charge decays in some finite volume is given by the surface integral of the space-like part $K_{i}$ of the Chern-Simons current (\ref{cs_current}) over the boundary of this volume. We can therefore interpret the non-conservation of the covariant current as an outflow of charge through the boundary. Whether or not such an outflow is possible depends on the physical setup in which one observes the Chiral Magnetic Effect. While in a closed system such an outflow would be impossible, it would be probably present for an expanding plasma or for a sample of Weyl semimetal attached to leads.

 We have pointed out that the vector-vector correlator which violates the vector Ward identities is obtained if one sums up the divergent contributions of different chiral states before integrating over the loop momentum. Within the Pauli-Villars regularization such summation leads in fact to the covariant form of the electric current. Thus in order to calculate the correlators of consistent currents, the contributions of each chiral sector should be individually regularized.

 In the context of condensed matter physics, our results imply that anomalous transport properties of lattice systems which have chiral fermions as their low-energy excitations are determined by the dispersion relation in the whole Brillouin zone and not just by small regions in the vicinity of the Fermi points/arcs. Another interesting observation is that in the case of chirally imbalanced matter at nonzero chiral chemical potential the Dirac mass plays quite an unusual role: instead of introducing a threshold for producing physical particles, it changes the exponential decay of the Fermi-Dirac distribution at large energies into a power-law decay. Such a change of a single-particle distribution might lead to some interesting non-Fermi-liquid behavior.

 We have considered several different regularizations of the anomalous vector-vector correlator such as the Pauli-Villars regularization and Wilson-Dirac and overlap fermions on the lattice. All of them agree up to some lattice artifacts. This suggests that a fully consistent chiral lattice regularization is in fact not so important for the observation of anomalous transport. It is only sufficient to require that a regularization reproduces the axial anomaly in the limit of infinite UV cutoff, which is the case for all the regularizations which we have considered. However, we have observed that the use of chiral lattice fermions significantly reduces finite-volume and finite-spacing effects. We have also found that once one uses lattice chiral fermions, the L\"uscher implementation of the chiral symmetry on the lattice becomes crucial. In particular, the anomalous vector-vector correlator decreases practically to zero once one uses ordinary chiral rotations instead of the L\"uscher transformations. It might be also interesting to consider anomalous transport for staggered fermions and see whether the contributions of the tastes of opposite chirality cancel in the chiral separation or the chiral magnetic conductivities.

 A novel observation which we make in this paper is that the asymptotic behavior of the anomalous current-current correlators in the limit of large momentum (much larger than the corresponding chemical potential) is also fixed by vector and axial Ward identities. The relation can be most easily illustrated by considering the derivatives of the axial-vector and vector-vector correlators over the chemical or the chiral chemical potentials, respectively. It is easy to see that these derivatives correspond to the fermion loop with insertion of two vector and one axial vertices and with zero momentum on one of the vertices. In other words, we have a process in which a virtual photon creates two virtual fermions, one of which interacts with an infinitely soft photon or axion. The fermions then annihilate and produce either an axion or a photon. If the momentum the initial virtual photon is sufficiently large, it does not notice the low-energy fermionic states within the Fermi surface, and our fermion loop effectively turns into the usual triangle diagram in the vacuum. For the kinematics which corresponds to the Chiral Separation Effect, this diagram is equal to zero. In the case of the Chiral Magnetic Effect, one picks up exactly the anomalous part of the triangle diagram if the chiral chemical potential (which corresponds to a soft axion within our analogy) slowly changes in time. In the case of spatially modulated chemical potential which is constant in time one obtains the transverse part of the triangle diagram. It can be still related to the anomaly by using the perturbative non-renormalization theorems of \cite{Knecht:04:1}, however, one can also expect some non-perturbative contributions in the phase with broken chiral symmetry. On the lattice, the latter limit of spatially modulated chiral chemical potential appears much more natural since it still corresponds to the equilibrium state of the system. From these arguments we conclude that the anomalous axial-vector correlator $\Pi_{12}^{AV}\lr{k_3}$ should vanish in the limit of large momentum $k_3$, and the anomalous vector-vector correlator $\Pi_{12}^{VV}\lr{k_3}$ should asymptotically approach the linear function $\Pi_{12}^{VV}\lr{k_3} = \frac{i k_3 \mu_A}{2 \pi^2}$. The slope of this function might in general be affected by non-perturbative effects and thus is an interesting quantity for a study on the lattice or within effective models. It is also worth noting that as the Dirac mass increases, this linear behavior is approached at higher momenta. Therefore in the limit of large mass the regularized chiral magnetic conductivity should vanish, as one could expect.

 These findings also agree with the results of holographic calculations \cite{Gynther:10:1}. This is to be expected, since the theorems of \cite{Knecht:04:1, Vainshtein:03:1, Vainshtein:03:2} are valid when the chiral symmetry is not broken, and the holographic model of \cite{Gynther:10:1} corresponds to the deconfinement phase with restored chiral symmetry. It might be also interesting to investigate possible non-perturbative corrections to $\Pi_{12}^{VV}\lr{k_3}$ in the phase with broken chiral symmetry.

 It is important to stress that while the behavior of anomalous current-current correlators (\ref{CSEPolarizationTensor}) and (\ref{CMEPolarizationTensor}) at small and large momenta is fixed by Ward identities, at intermediate momenta they would be, in general, changed due to interactions. For example, the radius of the Fermi surface might change as one goes from deconfinement to the confinement regime. In particular, it would be interesting to understand whether the ``chiral'' Fermi surface associated with finite chiral chemical potential still exists in the confinement regime.

\begin{acknowledgments}
 I am grateful to G.~Bali, F.~Bruckmann, G.~Endrody, U.~Gursoy, K.~Jensen, T.~Kalaydzhyan, K.~Landsteiner, A.~Sch\"{a}fer, O.~Teryaev, N.~Yamamoto and especially to A.~Sadofyev for many interesting discussions, comments and remarks which have stimulated and improved this work. This work was supported by the S.~Kowalewskaja award from the Alexander von Humboldt foundation.
\end{acknowledgments}

\appendix

\section{Correlator of axial and vector currents continuum free fermions at finite chemical potential}
\label{apdx:cse_continuum}

 In this Appendix we give some technical details of the calculation of the correlator of axial and vector currents (\ref{CSEPolTens}) at finite chemical potential $\mu_V$ for free Dirac fermions in the continuum.

 Our starting point is the expression (\ref{CSEPolTens1}) with $\mathcal{D}\lr{p, \mu_V} = i \gamma_{\mu} p_{\mu} + \mu_V \gamma_0 + m$ being the massive Dirac operator at finite chemical potential $\mu_V$. Evaluating the trace over spinor indices in (\ref{CSEPolTens1}) explicitly and using the identity $\tr\lr{\gamma_5 \gamma_{\alpha} \gamma_{\nu} \gamma_{\beta} \gamma_{\mu}} = -4 \epsilon_{\alpha \nu \beta \mu}$, we arrive at
\begin{eqnarray}
\label{CSEPolTens2Apdx}
 \Pi^{AV}_{\mu\nu}\lr{k} = \int \frac{d^4 l}{\lr{2 \pi}^4}
 \nonumber \\
 \frac{4 \epsilon_{\alpha \nu \beta \mu} k_{\alpha} l_{\beta}}{\lr{\lr{l+k/2}^2 + m^2} \lr{\lr{l - k/2}^2 + m^2}} .
\end{eqnarray}
As discussed in Subsection \ref{subsec:cse_continuum}, in order to take into account the chemical potential $\mu_V$,  in (\ref{CSEPolTens2Apdx}) we have to shift the contour of integration over $l_0$ as $l_0 \rightarrow l_0 - i \mu_V$.

In order to apply the Kubo relation (\ref{Kubo_CSE}), we set $\mu = 1$, $\nu = 2$ and assume that the only nonzero component of $k$ is $k_3$ (thus $k$ is purely space-like vector). In order to implement finite-temperature regularization, we separately integrate over the time-like component of the loop momentum $l_0$ and over the space-like components which we denote as $\vec{l}$:
\begin{widetext}
\begin{eqnarray}
\label{CSEPolTens3Apdx}
 \Pi^{AV}_{12}\lr{k_3} = -4 k_3 \int \frac{d^3 \vec{l}}{\lr{2 \pi}^3} \int \frac{d l_0}{2 \pi}
 \frac{l_0 - i \mu_V}{\lr{\lr{l_0 - i \mu_V}^2 + p^2 + m^2} \lr{\lr{l_0 - i \mu_V}^2 + q^2 + m^2}}
 = \nonumber \\ =
 -4 k_3 \int \frac{d^3 \vec{l}}{\lr{2 \pi}^3} \frac{1}{2\lr{p^2 - q^2}}
 \times \nonumber \\ \times
 \int \frac{d l_0}{2 \pi}
 \lr{
  \frac{1}{l_0 - i \mu_V - i \sqrt{q^2 + m^2}}
 +\frac{1}{l_0 - i \mu_V + i \sqrt{q^2 + m^2}}
 -\frac{1}{l_0 - i \mu_V - i \sqrt{p^2 + m^2}}
 -\frac{1}{l_0 - i \mu_V + i \sqrt{p^2 + m^2}}
 }
\end{eqnarray}
\end{widetext}
where we have denoted $p = |\vec{l} + \vec{k}/2|$, $q = |\vec{l} - \vec{k}/2|$. We now use the identity
\begin{eqnarray}
\label{MatsubaraIntegration}
 \int\limits_{-\infty}^{+\infty} \frac{d l_0}{2 \pi} \frac{1}{l_0 - i \epsilon} =
 \frac{i}{2} \, \sign\lr{\epsilon}
\end{eqnarray}
to integrate over $l_0$. The result is
\begin{widetext}
\begin{eqnarray}
\label{CSEPolTens4Apdx}
 \Pi^{AV}_{12}\lr{k_3} = 2 i k_3 \int \frac{d^3 \vec{l}}{\lr{2 \pi}^3} \frac{1}{4 k_3 l_3}
 \times \nonumber \\ \times
 \lr{
  \sign\lr{\mu_V + \sqrt{p^2 + m^2}}
 +\sign\lr{\mu_V - \sqrt{p^2 + m^2}}
 -\sign\lr{\mu_V + \sqrt{q^2 + m^2}}
 -\sign\lr{\mu_V - \sqrt{q^2 + m^2}}
 } ,
\end{eqnarray}
where we have replaced $2 \lr{p^2 - q^2} = 2 \lr{\vec{l} + \vec{k}/2}^2 - 2 \lr{\vec{l} - \vec{k}/2}^2 = 4 \vec{k} \cdot \vec{l} = 4 k_3 l_3$. For definiteness, let us now assume that the chemical potential $\mu_V$ and the external momentum $k_3$ are positive. The equation (\ref{CSEPolTens4Apdx}) then simplifies to
\begin{eqnarray}
\label{CSEPolTens5Apdx}
 \Pi^{AV}_{12}\lr{k_3} = i \int \frac{d^3 \vec{l}}{\lr{2 \pi}^3} \frac{1}{l_3}
  \lr{
 \theta\lr{\mu_V - \sqrt{\lr{\vec{l} + \vec{k}/2}^2 + m^2}} - \theta\lr{\mu_V - \sqrt{\lr{\vec{l} - \vec{k}/2}^2 + m^2}}
 }
 = \nonumber \\ =
 i \int\limits_{-\infty}^{+\infty} \frac{d l_3}{2 \pi l_3} \int\limits_{0}^{+\infty}
  \frac{d\lr{\pi l_{\perp}^2}}{\lr{2 \pi}^2}
  \lr{
 \theta\lr{\mu_V - \sqrt{\lr{l_3 + k_3/2}^2 + l_{\perp}^2 + m^2}} - \theta\lr{\mu_V - \sqrt{\lr{l_3 - k_3/2}^2 + l_{\perp}^2 + m^2}} } ,
\end{eqnarray}
where we have denoted $l_{\perp} = \sqrt{l_1^2 + l_2^2}$.
\end{widetext}

 From the expressions above we see that the integration in the momentum space is restricted to two disjoint regions in momentum space which are depicted on Fig. \ref{fig:finite_mu_regions}. If the external momentum $|\vec{k}| > 2 \sqrt{\mu_V^2 - m^2}$, these regions are just two spheres of radii $\sqrt{\mu_V^2 - m^2}$ centered around $l_3 = \pm k_3/2$, $l_{1,2} = 0$. For $|k_3| < 2 \sqrt{\mu_V^2 - m^2}$, the intersection of these two spheres is removed from the integral. Let us now proceed with the calculation of the polarisation tensor (\ref{CSEPolTens5Apdx}) by integrating out $l_{\perp}$. In the case of large external momentum $k_3 > 2 \mu_V$ integration over $l_{\perp}$ yields the area of the section of the sphere by the plane $l_3 = \const$, and we obtain:
\begin{eqnarray}
\label{CSEPolTensLargeK3IntegrandApdx}
 \Pi^{AV}_{12}\lr{k_3} = -2 i \int\limits_{k_3/2 - \sqrt{\mu_V^2 - m^2}}^{k_3/2 + \sqrt{\mu_V^2 - m^2}} \frac{d l_3}{2 \pi l_3}
  \times \nonumber \\ \times
  \frac{\pi \lr{\mu_V^2 - m^2 - \lr{l_3 - k_3/2}^2}}{\lr{2 \pi}^2} ,
\end{eqnarray}
where the factor of two in front of the r.h.s. takes into account the contribution of the two spheres in momentum space, both the one centered at $l_3 = -k_3/2$ and the one at $l_3 = k_3/2$. Integration over $l_3$ can be performed analytically, and we obtain
\begin{eqnarray}
\label{CSEPolTensLargeK3Apdx}
 \Pi^{AV}_{12}\lr{k_3} = -\frac{i}{\lr{2 \pi}^2}
 \left( \sqrt{\mu_V^2 - m^2} k_3
 - \right. \nonumber \\ \left. -
 \lr{\mu_V^2 - m^2 - k_3^2/4} \, \log\lr{\frac{k_3 - 2 \sqrt{\mu_V^2 - m^2}}{k_3 + 2 \sqrt{\mu_V^2 - m^2}}}
 \right) ,
\end{eqnarray}

 Let us now turn to the case of small momenta $k_3 < 2 \sqrt{\mu_V^2 - m^2}$, when the two spheres intersect. The intersection region has to be explicitly subtracted from integration, which leads to
\begin{widetext}
\begin{eqnarray}
\label{CSEPolTensSmallK3IntegrandApdx}
 \Pi^{AV}_{12}\lr{k_3} =
   -2 i \int\limits_{\sqrt{\mu_V^2 - m^2} - k_3/2}^{\sqrt{\mu_V^2 - m^2} + k_3/2} \frac{d l_3}{2 \pi l_3}
   \frac{\pi \lr{\mu_V^2 - m^2 - \lr{l_3 - k_3/2}^2}}{\lr{2 \pi}^2}
 - \nonumber \\ -
   2 i \int\limits_{0}^{\sqrt{\mu_V^2 - m^2} - k_3/2} \frac{d l_3}{2 \pi l_3}
   \frac{\pi \lr{\mu_V^2 - m^2 - \lr{l_3 - k_3/2}^2} - \pi \lr{\mu_V^2 - m^2 - \lr{l_3 + k_3/2}^2}}{\lr{2 \pi}^2}.
\end{eqnarray}
The integration over $l_3$ can be again performed analytically, and we get
\begin{eqnarray}
\label{CSEPolTensSmallK3Apdx}
 \Pi^{AV}_{12}\lr{k_3} =
 -\frac{2 i k_3}{\lr{2 \pi}^2} \lr{\sqrt{\mu_V^2 - m^2} - k_3/2}
 - \nonumber \\ -
 \frac{i k_3}{\lr{2 \pi}^2} \lr{k_3 - \sqrt{\mu_V^2 - m^2}}
 +
 \frac{i \lr{\mu_V^2 - m^2 - k_3^2/4}}{\lr{2 \pi}^2} \log\lr{\frac{2 \sqrt{\mu_V^2 - m^2} - k_3}{2 \sqrt{\mu_V^2 - m^2} + k_3}}
 = \nonumber \\ =
 -\frac{i}{\lr{2 \pi}^2}
 \lr{ \sqrt{\mu_V^2 - m^2} k_3
 - \lr{\mu_V^2 - m^2 - k_3^2/4} \, \log\lr{\frac{2 \sqrt{\mu_V^2 - m^2} - k_3}{2 \sqrt{\mu_V^2 - m^2} + k_3}}
 }
\end{eqnarray}
\end{widetext}
We see that both results (\ref{CSEPolTensLargeK3Apdx}) and (\ref{CSEPolTensSmallK3Apdx}) can be unified in a single expression (\ref{CSEPolTens}).

\section{Correlator of two vector currents for continuum free fermions at finite chiral chemical potential}
\label{apdx:cme_continuum}

 In this Appendix we summarize the details of the calculation of the correlator (\ref{CMEPolarizationTensor}) of two vector currents for free Dirac fermions.

 We start with the expression (\ref{CMEPolTens1}) with $\mathcal{D}\lr{p, \mu_A} = i \gamma_{\mu} p_{\mu} + \mu_A \gamma_0 \gamma_5 + m$ being the massive Dirac operator at chiral chemical potential $\mu_A$. Representing $\mathcal{D}\lr{p, \mu_A}$ in the chiral block form and performing the block matrix inversion, we can write the Dirac propagator $\mathcal{D}^{-1}\lr{p, \mu_A}$ as
\begin{eqnarray}
\label{propagator_finite_mu5Apdx}
 \mathcal{D}^{-1}\lr{p, \mu_A} =
 \frac{1}{p_0^2 + \lr{\dslash{p} - \mu_A}^2 + m^2}
 \times \nonumber \\ \times
 \left(
   \begin{array}{cc}
                                m & -i p_0 - \dslash{p} + \mu_A \\
      -i p_0 + \dslash{p} - \mu_A &                           m \\
   \end{array}
 \right)  ,
\end{eqnarray}
where $\dslash{p} = \sigma_k p_k$, $p_k$ are spatial components of the momentum and $\sigma_k$ with $k = 1 \ldots 3$ are the Pauli matrices. It is convenient to represent the operator $\lr{p_0^2 + \lr{\dslash{p} - \mu_A}^2 + m^2}^{-1}$ in (\ref{propagator_finite_mu5Apdx}) in terms of the chiral projectors $\mathcal{P}_{\pm} = \frac{1 \pm \dslash{p}/|\vec{p}|}{2}$ as
\begin{eqnarray}
\label{propagator_mu5_chir_decomp1Apdx}
 \frac{1}{p_0^2 + \lr{\dslash{p} - \mu_A}^2 + m^2} = \sum\limits_{s = \pm} \mathcal{P}_{s}\lr{\vec{p}} G_{s}\lr{p_{\mu}, \mu_A}  ,
 \nonumber \\
 G_{s}\lr{p_{\mu}, \mu_A} = \frac{1}{p_0^2 + \lr{|\vec{p}| - s \mu_A}^2 + m^2}  ,
\end{eqnarray}
where $s = \pm$ labels chiral states which saturate the propagator (\ref{propagator_mu5_chir_decomp1Apdx}), namely, the states for which the spin is parallel or antiparallel to the spatial momentum. We now use the identity $\mathcal{P}_s \, \dslash{p} = s |\vec{p}| \mathcal{P}_s$ and obtain for the propagator (\ref{propagator_mu5_chir_decomp1Apdx}):
\begin{eqnarray}
\label{propagator_mu5_chir_decomp2Apdx}
 \mathcal{D}^{-1}\lr{p_{\mu}, \mu_A}
  =
 \sum\limits_{s} G_s\lr{p_{\mu}, \mu_A} \mathcal{P}_s
 \times \nonumber \\ \times
 \left(
   \begin{array}{cc}
                                  m & -i p_0 + \mu_A - s |\vec{p}| \\
       -i p_0 - \mu_A + s |\vec{p}| &                            m \\
   \end{array}
 \right)  ,
\end{eqnarray}
where the matrix elements in the second line are just proportional to two by two identity matrices. We now consider only the space-like components of the polarisation tensor (\ref{CMEPolarizationTensor}), which enter the Kubo formulae (\ref{Kubo_CME}). Spatial Dirac gamma-matrices can be written in chiral block form as
\begin{eqnarray}
\label{gamma_matr_as_dir_prod}
 \gamma_k = \sigma_k \, \left(
   \begin{array}{cc}
     0 & -i \\
     i &  0 \\
   \end{array}
 \right)  ,
\end{eqnarray}
where again the elements of the second matrix are proportional to $2 \times 2$ identity matrices (one could also write a direct product operator between the two matrices in the above equation).

 Inserting the expressions (\ref{propagator_mu5_chir_decomp2Apdx}) and (\ref{gamma_matr_as_dir_prod}) into (\ref{CMEPolTens1}) and taking into account that the trace of the direct product of matrices is a product of traces, we get
\begin{widetext}
\begin{eqnarray}
\label{CMEPolTens2Apdx}
  \Pi^{VV}_{kl}\lr{k} = \int \frac{d^4 l}{\lr{2 \pi}^4}
  \sum\limits_{s,s'}
  G_{s}\lr{l_{\mu} + k_{\mu}/2} G_{s'}\lr{l_{\mu} - k_{\mu}/2}
  \tr\lr{\sigma_k \mathcal{P}_s\lr{\vec{l} + \vec{k}/2} \sigma_l \mathcal{P}_{s'}\lr{\vec{l} - \vec{k}/2} }
  \times \nonumber \\ \times
  \tr\lr{
  \left(
   \begin{array}{cc}
    0 & -i \\
    i &  0 \\
   \end{array}
 \right)
  \left(
   \begin{array}{cc}
                          m & -i l_0 + \mu_A - s p \\
       -i l_0 - \mu_A + s p &                    m \\
   \end{array}
 \right)
 \left(
   \begin{array}{cc}
    0 & -i \\
    i &  0 \\
   \end{array}
 \right)
 \left(
   \begin{array}{cc}
                           m & -i l_0 + \mu_A - s' q \\
       -i l_0 - \mu_A + s' q &                     m \\
   \end{array}
 \right) } ,
\end{eqnarray}
\end{widetext}
where we have denoted $p = |\vec{l} + \vec{k}/2|$ and $q = |\vec{l} - \vec{k}/2|$. Lets evaluate first the trace over Weyl indices in the first line of (\ref{CMEPolTens2Apdx}). For the sake of simplicity we also denote $\vec{p} = \vec{l} + \vec{k}/2$ and $\vec{q} = \vec{l} - \vec{k}/2$. We then obtain
\begin{eqnarray}
\label{CMEPolTensTrace1.1Apdx}
 \tr\lr{\sigma_k \mathcal{P}_s\lr{\vec{p}} \sigma_l \mathcal{P}_{s'}\lr{\vec{q}}}
 = \nonumber \\ =
 \frac{\delta_{kl}}{2}
 +
 \frac{s'}{4 q} \tr\lr{\sigma_l \dslash{q} \sigma_k}
 +
 \frac{s}{4 p} \tr\lr{\sigma_k \dslash{p} \sigma_l}
 + \nonumber \\ +
 + \frac{s s'}{4 p q} \tr\lr{\sigma_k \dslash{p} \sigma_l \dslash{q}}
 = \nonumber \\ =
 \frac{\delta_{kl}}{2}
 +
 \frac{s'}{4 q} \, 2 i \epsilon_{kli} q_i
 +
 \frac{s}{4 p} \, 2 i \epsilon_{lki} p_i
 + \nonumber \\ +
 \frac{s s'}{4 p q} \lr{2 p_k q_l + 2 p_l q_k - 2 \delta_{kl} \lr{\vec{p} \cdot \vec{q}}}  .
\end{eqnarray}
In order to get the chiral magnetic conductivity from the Kubo relation (\ref{Kubo_CME}), we only need to calculate the above expression for $k = 1$, $l = 2$ and with $k_3$ being the only nonzero component of $k$. For such kinematics we get a simple result
\begin{eqnarray}
\label{CMEPolTensTrace1.2Apdx}
 \tr\lr{\sigma_1 \mathcal{P}_s\lr{\vec{p}} \sigma_2 \mathcal{P}_{s'}\lr{\vec{q}}}
 = \nonumber \\ =
 -\frac{i}{2} \lr{s \frac{p_3}{p} - s' \frac{q_3}{q}} + \frac{s s' l_1 l_2}{p q}  .
\end{eqnarray}
The last summand in the above expression will not contribute to the Chiral Magnetic Effect in any reasonable regularization by virtue of rotational invariance in the $1,2$ plane, thus we discard it in what follows.

 A direct calculation of the trace in the second line of (\ref{CMEPolTens2Apdx}) yields
\begin{eqnarray}
\label{CMEPolTensTrace2.1Apdx}
 2 \lr{m^2 + l_0^2 - \lr{\mu_A - s p} \lr{\mu_A - s' q} } .
\end{eqnarray}
The current-current correlator (\ref{CMEPolTens2Apdx}) can be now written as
\begin{eqnarray}
\label{CMEPolTens3Apdx}
 \Pi^{VV}_{12}\lr{k_3} = -i \int \frac{d^3 l}{\lr{2 \pi}^3}\sum\limits_{s,s'} \, \lr{s \frac{p_3}{p} - s' \frac{q_3}{q}} \int \frac{d l_0}{2 \pi}
  \nonumber \\
  \frac{m^2 + l_0^2 - s s' \lr{p - s \mu_A} \lr{q - s' \mu_A} }{
  \lr{l_0^2 + \lr{p - s \mu_A}^2 + m^2}
  \lr{l_0^2 + \lr{q - s' \mu_A}^2 + m^2}
  } .
\end{eqnarray}
We first take the integral over $l_0$. Some simple algebraic manipulations lead to
\begin{eqnarray}
\label{CMEPolTens4Apdx}
\int \frac{d l_0}{2 \pi}
  \frac{m^2 + l_0^2 - s s' \epsilon_p \epsilon_q }{
  \lr{l_0^2 + \epsilon_p^2 + m^2}
  \lr{l_0^2 + \epsilon_q^2 + m^2} }
  =
  \frac{1}{\epsilon_p - s s' \epsilon_q}
  \times \nonumber \\ \times
  \int \frac{d l_0}{2 \pi}
  \lr{
  \frac{\epsilon_p}{l_0^2 + \epsilon_p^2 + m^2}
  -
  \frac{s s' \epsilon_q}{l_0^2 + \epsilon_q^2 + m^2}
  }
  = \nonumber \\ =
  \frac{1}{2 \lr{\epsilon_p - s s' \epsilon_q}}
  \lr{\frac{\epsilon_p}{\sqrt{\epsilon_p^2 + m^2}} - \frac{s s' \epsilon_q}{\sqrt{\epsilon_q^2 + m^2}}}  ,
\end{eqnarray}
where we have denoted $\epsilon_p = p - s \mu_A$, $\epsilon_q = q - s' \mu_A$. Combining the expressions (\ref{CMEPolTens4Apdx}) and (\ref{CMEPolTens3Apdx}), we finally obtain the representation (\ref{CMEPolTens2}) of the polarization tensor (\ref{CMEPolarizationTensor}) in terms of spatial loop momentum and the chiral states labeled by $s$ and $s'$.

 Following \cite{Kharzeev:09:1}, we now sum up the contributions of chiral states with different $s$ and $s'$ to the current-current correlator (\ref{CMEPolTens2}) before integrating over spatial loop momentum:
\begin{widetext}
\begin{eqnarray}
\label{CMEPolTens5Apdx}
 \Pi^{VV}_{12}\lr{k_3} = -i \int \frac{d^3 l}{\lr{2 \pi}^3}
 \left(
 \frac{q p_3 - p q_3}{2 p q \lr{p - q}}
 \lr{\frac{p - \mu_A}{\sqrt{\lr{p - \mu_A}^2 + m^2}} - \frac{q - \mu_A}{\sqrt{\lr{q - \mu_A}^2 + m^2}}}
 + \right. \nonumber \\ +
 \frac{q p_3 + p q_3}{2 p q \lr{p + q}}
 \lr{\frac{p - \mu_A}{\sqrt{\lr{p - \mu_A}^2 + m^2}} + \frac{q + \mu_A}{\sqrt{\lr{q + \mu_A}^2 + m^2}}}
 + \nonumber \\ +
 \frac{ - q p_3 - p q_3}{2 p q \lr{p + q}}
 \lr{\frac{p + \mu_A}{\sqrt{\lr{p + \mu_A}^2 + m^2}} + \frac{q - \mu_A}{\sqrt{\lr{q - \mu_A}^2 + m^2}}}
 -\nonumber \\ \left. -
 \frac{q p_3 - p q_3}{2 p q \lr{p - q}}
 \lr{\frac{p + \mu_A}{\sqrt{\lr{p + \mu_A}^2 + m^2}} - \frac{q + \mu_A}{\sqrt{\lr{q + \mu_A}^2 + m^2}}}
 \right)
 = \nonumber \\ =
 -i \int \frac{d^3 l}{\lr{2 \pi}^3}
 \lr{\frac{q p_3 + p q_3}{2 p q \lr{p + q}} + \frac{q p_3 - p q_3}{2 p q \lr{p - q}}}
 \times \nonumber \\ \times
 \lr{
 \frac{p - \mu_A}{\sqrt{\lr{p - \mu_A}^2 + m^2}}
 -
 \frac{q - \mu_A}{\sqrt{\lr{q - \mu_A}^2 + m^2}}
 +
 \frac{q + \mu_A}{\sqrt{\lr{q + \mu_A}^2 + m^2}}
 -
 \frac{p + \mu_A}{\sqrt{\lr{p + \mu_A}^2 + m^2}}
 }
\end{eqnarray}
\end{widetext}
A direct calculation shows that
\begin{eqnarray}
\label{CMEPolTensSimpl1Apdx}
 \lr{\frac{q p_3 + p q_3}{2 p q \lr{p + q}} + \frac{q p_3 - p q_3}{2 p q \lr{p - q}}} = \frac{1}{2 l_3}  ,
\end{eqnarray}
which together with (\ref{CMEPolTens5Apdx}) leads to the expression (\ref{CMEPolTens3}).

 In order to apply the Pauli-Villars regularization to the current-current correlator (\ref{CMEPolarizationTensor}), we also consider the expression (\ref{CMEPolTens3}) in the limit of infinitely large quark mass $m$. In this limit we can approximate the summands in the parentheses in (\ref{CMEPolTens3}) as
\begin{eqnarray}
\label{CMEPolTensSimplHeavyApdx}
 \frac{p - \mu_A}{\sqrt{\lr{p - \mu_A}^2 + m^2}}
 -
 \frac{p + \mu_A}{\sqrt{\lr{p + \mu_A}^2 + m^2}}
 = \nonumber \\ =
 -2 \mu_A \frac{\partial}{\partial p} \frac{p}{\sqrt{p^2 + m^2}}  + O\lr{\frac{\mu_A^3}{m^3}}
 = \nonumber \\ =
 -2 \mu_A \frac{m^2}{\lr{m^2 + p^2}^{3/2}} + O\lr{\frac{\mu_A^3}{m^3}},
\end{eqnarray}
and similarly also for summands which involve $q \pm \mu_A$. We then arrive at
\begin{widetext}
\begin{eqnarray}
\label{CMEPolTensHeavy1Apdx}
 \Pi^{VV}_{12}\lr{k_3}
 =
 i \mu_A \int \frac{d^3 l}{\lr{2 \pi}^3} \frac{1}{l_3}
 \lr{\frac{m^2}{\lr{m^2 + p^2}^{3/2}} - \frac{m^2}{\lr{m^2 + q^2}^{3/2}}}
 = \nonumber \\ =
 i \mu_A \int\limits_{-\infty}^{+\infty} \frac{d l_3}{2 \pi l_3}
 \int\limits_{0}^{+\infty} \frac{d \lr{\pi l_{\perp}^2}}{\lr{2 \pi}^2}
 \lr{\frac{m^2}{\lr{m^2 + \lr{l_3 + k_3/2}^2 + l_{\perp}^2}^{3/2}} - \frac{m^2}{\lr{m^2 + \lr{l_3 - k_3/2}^2 + l_{\perp}^2}^{3/2}}}
 = \nonumber \\ =
 2 i \mu_A \int\limits_{-\infty}^{+\infty} \frac{d l_3}{2 \pi l_3}
 \frac{\pi m^2}{\lr{2 \pi}^2}
 \lr{\frac{1}{\sqrt{m^2 + \lr{l_3 + k_3/2}^2}} - \frac{1}{\sqrt{m^2 + \lr{l_3 - k_3/2}^2}}}  ,
\end{eqnarray}
\end{widetext}
where in the last line we have performed integration over $l_{\perp}$. Again, in the limit of very large mass $m \gg k_3$ we can approximate
\begin{eqnarray}
\label{CMEPolTensSimpl3Apdx}
 \frac{1}{\sqrt{m^2 + \lr{l_3 + k_3/2}^2}} - \frac{1}{\sqrt{m^2 + \lr{l_3 - k_3/2}^2}}
 = \nonumber \\ =
 k_3 \frac{\partial}{\partial l_3} \frac{1}{\sqrt{m^2 + l_3^2}} + O\lr{\frac{k_3^3}{m^4}}
 = \nonumber \\ =
 - \frac{k_3 l_3}{\lr{m^2 + l_3^2}^{3/2}}  + O\lr{\frac{k_3^3}{m^4}} .
\end{eqnarray}
Taking into account the above expression, we obtain
\begin{eqnarray}
\label{CMEPolTensHeavy2Apdx}
 \Pi^{VV}_{12}\lr{k_3}
 =
 - \frac{i \mu_A k_3}{\lr{2 \pi}^2} \int\limits_{-\infty}^{+\infty}
 \frac{d l_3 \, m^2}{\lr{m^2 + l_3^2}^{3/2}}
 = \nonumber \\ =
 - \frac{i \mu_A k_3}{2 \pi^2} ,
\end{eqnarray}
that is, the result which depends linearly on both $\mu_A$ and $k_3$ as long as $\mu_A \ll m$ and $k_3 \ll m$.

\section{Derivatives of the overlap Dirac operator over vector and axial gauge fields}
\label{apdx:overlap_derivatives}

 In this Appendix we construct the derivatives of the overlap Dirac operators (\ref{overlap_finite_mu}) and (\ref{overlap_mu5_implicit}) at finite chemical or chiral chemical potentials with respect to the vector and axial gauge fields. Such derivatives are necessary for the calculation of the axial-vector and vector-vector correlators (\ref{av_correlator_observable}) and (\ref{vv_correlator_observable}). Our aim is to express these derivatives in terms of the derivatives of the local operator $\mathcal{H} = \gamma_5 \mathcal{D}_w$ which enters the overlap definition (\ref{overlap_finite_mu}) and the derivatives of its eigenvalues (\ref{g5dw_eigval}) and left/right eigenvectors (\ref{g5dw_eigenvecs}). Since the overlap Dirac operator at finite chiral chemical potential (\ref{overlap_mu5_implicit}) is also defined in terms of the overlap at finite chemical potential $\mu_V$, the derivatives of the former can also be expressed in terms of derivatives of $\mathcal{H}$.

 To this end let us first express the derivatives of the eigenvectors and eigenvalues of $\mathcal{H}$ with respect to the vector gauge field $V_{x, \mu}$ in terms of derivatives of $\mathcal{H}$. By differentiating the equations (\ref{lr_decomposition_general}) it is easy to arrive at the following relations:
\begin{eqnarray}
\label{eval_deriv}
 \partial^{V}_{x,\mu} \lambda_i = \bra{L_i} \lr{\partial^{V}_{x,\mu} \mathcal{H}} \ket{R_i}
\\
\label{evec_deriv}
 \bra{L_j} \partial^{V}_{x,\mu} \ket{R_i} = \frac{\bra{L_j}  \partial^{V}_{x,\mu} \mathcal{H} \ket{R_i}}{\lambda_i - \lambda_j}, \quad i \neq j .
\end{eqnarray}
The relations (\ref{evec_deriv}) alone are not enough to completely fix the derivatives of the eigenvectors $\bra{L_i}$ and $\ket{R_i}$. The reason is that there is still an ambiguity in the equations (\ref{lr_decomposition_general}). Namely, one can redefine $\ket{R_i} \rightarrow e^{\Phi_i} \ket{R_i}$ and $\bra{L_i} \rightarrow \bra{L_i} e^{-\Phi_i}$, with $\Phi_i$ being arbitrary complex numbers which can have arbitrary dependance on gauge fields. In order to fix this ambiguity, it is convenient to impose an additional condition $\bra{L_i} \partial^{V}_{x,\mu} \ket{R_i} = 0$. With this additional relation, the derivatives of eigenvectors can be written as
\begin{eqnarray}
\label{revec_deriv_full}
 \partial^V_{x, \mu} \ket{R_i} = \sum\limits_{j \neq i}
 \frac{\ket{R_j}\bra{L_j} \partial^V_{x, \mu} \mathcal{H} \ket{R_i}}{\lambda_i - \lambda_j}  ,
\\
\label{levec_deriv_full}
 \partial^V_{x, \mu} \bra{L_i} = \sum\limits_{j \neq i}
 \frac{\bra{L_i} \partial^V_{x, \mu} \mathcal{H} \ket{R_j}\bra{L_j}}{\lambda_i - \lambda_j}  .
\end{eqnarray}

 The derivative of the sign function $\sign\lr{\mathcal{H}}$ of the operator $\mathcal{H}$ over the vector gauge field $V_{x, \mu}$ can be now written as
\begin{eqnarray}
\label{overlap_derivative_vector1}
 \partial^V_{x, \mu} \, \sign\lr{\mathcal{H}}
 = \sum\limits_i \ket{R_i} \partial^V_{x, \mu} s_i \bra{L_i}
 + \nonumber \\ +
 \sum\limits_i \lr{\partial^V_{x, \mu} \ket{R_i} s_i \bra{L_i} + \ket{R_i} s_i \partial^V_{x, \mu} \bra{L_i}}  ,
\end{eqnarray}
where we have denoted $s_i \equiv \sign\lr{\re \lambda_i}$ for the sake of brevity. The term on the right-hand side in the first line contributes only when one of the eigenvalues $\lambda_i$ crosses the imaginary axis $\re \lambda_i = 0$. We have found that this never happens in practice, so one can safely omit this term. Using the relations (\ref{revec_deriv_full}) and (\ref{levec_deriv_full}), we then arrive at the following representation of the derivative of the overlap Dirac operator (\ref{overlap_finite_mu}) over $V_{x, \mu}$:
\begin{eqnarray}
\label{overlap_derivative_vector2}
 \partial^V_{x, \mu} \, \mathcal{D}_{ov}
 = \nonumber \\ =
 \sum\limits_{i \neq j}
 \frac{\gamma_5 \ket{R_i} \bra{L_i}  \partial^V_{x, \mu} \mathcal{H} \ket{R_j} \bra{L_j} \, \lr{s_i - s_j}}{\lambda_i - \lambda_j}  .
\end{eqnarray}
The derivative of $\mathcal{D}_{ov}$ over the axial gauge field $A_{x, \mu}$ can be expressed in terms of the above derivative over the vector gauge field using the relation (\ref{kikukawa_equation}) (for details see \cite{Kikukawa:98:1}). Taking into account that
\begin{eqnarray}
\label{modified_gamma5}
 \gamma_5 \lr{1 - \mathcal{D}_{ov}} = -\sign\lr{\gamma_5 \mathcal{D}_{w}\lr{\mu_V}}
= \nonumber \\ =
-\sum\limits_{i} \ket{R_i} s_i \bra{L_i}
\end{eqnarray}
we obtain from (\ref{overlap_derivative_vector2}):
\begin{eqnarray}
\label{overlap_derivative_axial}
 \partial^A_{x, \mu} \mathcal{D}_{ov}
 = \nonumber \\ =
 \sum\limits_{i \neq j}
 \frac{\gamma_5 \ket{R_i} \bra{L_i}  \partial^V_{x, \mu} \mathcal{H} \ket{R_j} \bra{L_j} \, \lr{1 - s_i s_j}}{\lambda_i - \lambda_j}  .
\end{eqnarray}

 In order to calculate the vector-axial correlator (\ref{av_correlator_observable}) we also need the mixed derivative $\partial^V_{y, \nu} \partial^A_{x, \mu} \mathcal{D}_{ov}$. To this end we differente the expression (\ref{overlap_derivative_axial}) over $V_{y, \nu}$, apply the chain rule and use once again the expressions (\ref{revec_deriv_full}), (\ref{levec_deriv_full}) and (\ref{eval_deriv}) for the derivatives of the eigenvectors and eigenvalues of $\mathcal{H}$. These manipulations lead to the following expression for $\partial^V_{y, \nu} \partial^A_{x, \mu} \mathcal{D}_{ov}$:
\begin{widetext}
\begin{eqnarray}
\label{overlap_derivative_vector_axial}
 \partial^V_{y, \nu} \, \partial^A_{x, \mu} \, \mathcal{D}_{ov}
 = 
 \sum\limits_{i \neq j, i \neq k}
 \frac{\gamma_5 \, \ket{R_k} \bra{L_k}  \partial^V_{y, \nu} \mathcal{H} \ket{R_i} \bra{L_i} \partial^V_{x, \mu} \mathcal{H} \ket{R_j} \bra{L_j} \, \lr{1 - s_i s_j}}{\lr{\lambda_i - \lambda_k} \, \lr{\lambda_i - \lambda_j}}
 + \nonumber \\ +
 \sum\limits_{i \neq j, k \neq j}
 \frac{\gamma_5 \, \ket{R_i} \bra{L_i}  \partial^V_{x, \mu} \mathcal{H} \ket{R_j} \bra{L_j} \partial^V_{y, \nu} \mathcal{H} \ket{R_k} \bra{L_k} \, \lr{1 - s_i s_j}}{\lr{\lambda_j - \lambda_k} \, \lr{\lambda_i - \lambda_j}}
+ \nonumber \\ +
 \sum\limits_{i \neq k, i \neq j}
 \frac{\gamma_5 \, \ket{R_i} \bra{L_i}  \partial^V_{y, \nu} \mathcal{H} \ket{R_k} \bra{L_k} \partial^V_{x, \mu} \mathcal{H} \ket{R_j} \bra{L_j} \, \lr{1 - s_i s_j}}{\lr{\lambda_i - \lambda_k} \, \lr{\lambda_i - \lambda_j}}
+ \nonumber \\ +
 \sum\limits_{j \neq k, j \neq i}
 \frac{\gamma_5 \, \ket{R_i} \bra{L_i}  \partial^V_{x, \mu} \mathcal{H} \ket{R_k} \bra{L_k} \partial^V_{y, \nu} \mathcal{H} \ket{R_j} \bra{L_j} \, \lr{1 - s_i s_j}}{\lr{\lambda_i - \lambda_j} \, \lr{\lambda_j - \lambda_k}}
+ \nonumber \\ +
 \sum\limits_{i \neq j} \frac{\gamma_5 \,  \ket{R_i} \bra{L_i} \partial^V_{x, \mu} \mathcal{H} \ket{R_j} \bra{L_j} \partial^V_{y, \nu} \mathcal{H} \ket{R_j} \bra{L_j} \, \lr{1 - s_i s_j}}{\lr{\lambda_i - \lambda_j}^2}
- \nonumber \\ -
 \sum\limits_{i \neq j} \frac{\gamma_5 \, \ket{R_i} \bra{L_i} \partial^V_{y, \nu} \mathcal{H} \ket{R_i} \bra{L_i} \partial^V_{x, \mu} \mathcal{H} \ket{R_j} \bra{L_j} \, \lr{1 - s_i s_j}}{\lr{\lambda_i - \lambda_j}^2}
+ \nonumber \\ +
 \sum\limits_{i \neq j} \frac{\gamma_5 \, \ket{R_i} \bra{L_i} \partial^V_{y, \nu} \partial^V_{x, \mu} \mathcal{H} \ket{R_j} \bra{L_j} \, \lr{1 - s_i s_j} }{\lambda_i - \lambda_j}  .
\end{eqnarray}
\end{widetext}
Here the first four summands come from the derivatives of the eigenvectors in (\ref{overlap_derivative_vector2}), the fifth and the sixth summands come from the derivative of the denominator in (\ref{overlap_derivative_vector2}) and the last summand comes from the second derivative of $\mathcal{H}$ itself.

 We now turn to the calculation of the derivatives of the overlap Dirac operator at finite chiral chemical potential, which is implicitly defined by the equation (\ref{overlap_mu5_implicit}). We first have to express the derivatives of $\mathcal{D}_{ov}\lr{\mu_A}$ in terms of the derivatives of $\tilde{\mathcal{D}}_{ov}\lr{\mu_A}$ given by (\ref{gw_projection}). Direct differentiation of the inverse Ginsparg-Wilson projection (\ref{inverse_gw_projection}) yields (for the sake of brevity we omit the argument $\mu_A$ of $\mathcal{D}_{ov}$ and $\tilde{\mathcal{D}}_{ov}$):
\begin{eqnarray}
\label{inverse_gw_proj_1st_deriv}
 \partial^V_{x, \mu} \mathcal{D}_{ov}
 = 
 \frac{2}{2 + \tilde{\mathcal{D}}_{ov}}
 \partial^V_{x, \mu} \tilde{\mathcal{D}}_{ov}
 \frac{2}{2 + \tilde{\mathcal{D}}_{ov}} ,
 \\
 \label{inverse_gw_proj_2nd_deriv}
 \partial^V_{y, \nu} \partial^V_{x, \mu} \mathcal{D}_{ov}
 = 
 \frac{2}{2 + \tilde{\mathcal{D}}_{ov}}
 \partial^V_{y, \nu} \partial^V_{x, \mu} \tilde{\mathcal{D}}_{ov}
 \frac{2}{2 + \tilde{\mathcal{D}}_{ov}}
 - \nonumber \\ -
 \frac{2}{2 + \tilde{\mathcal{D}}_{ov}} \partial^V_{x, \mu}\tilde{\mathcal{D}}_{ov}
 \frac{1}{2 + \tilde{\mathcal{D}}_{ov}} \partial^V_{y, \nu}\tilde{\mathcal{D}}_{ov}
 \frac{2}{2 + \tilde{\mathcal{D}}_{ov}}
 - \nonumber \\ -
 \frac{2}{2 + \tilde{\mathcal{D}}_{ov}} \partial^V_{y, \nu}\tilde{\mathcal{D}}_{ov}
 \frac{1}{2 + \tilde{\mathcal{D}}_{ov}} \partial^V_{x, \mu}\tilde{\mathcal{D}}_{ov}
 \frac{2}{2 + \tilde{\mathcal{D}}_{ov}}
\end{eqnarray}
Since equation (\ref{overlap_mu5_implicit}) is linear, the derivatives $\partial^V_{x, \mu}\tilde{\mathcal{D}}_{ov}\lr{\mu_A}$ and $\partial^V_{y, \nu} \partial^V_{x, \mu} \tilde{\mathcal{D}}_{ov}\lr{\mu_A}$ can be expressed in terms of the derivatives of the projected overlap Dirac operator at finite chemical potential $\tilde{\mathcal{D}}_{ov}\lr{\mu_V = \pm\mu_A}$ in a straightforward way.

 The next step is then to express the derivatives of the projected overlap operator $\tilde{\mathcal{D}}_{ov}\lr{\mu_V}$ in terms of the derivatives of the original overlap operator $\mathcal{D}_{ov}\lr{\mu_V}$. Explicit differentiation of (\ref{gw_projection}) yields the following expressions for these derivatives (we again omit the argument $\mu_V$ of $\mathcal{D}_{ov}$ and $\tilde{\mathcal{D}}_{ov}$):
\begin{eqnarray}
\label{gw_proj_1st_deriv}
 \partial^V_{x, \mu} \tilde{\mathcal{D}}_{ov}
 = 
 \frac{2}{2 - \mathcal{D}_{ov}}
 \partial^V_{x, \mu} \mathcal{D}_{ov}
 \frac{2}{2 - \mathcal{D}_{ov}} ,
 \\
 \label{gw_proj_2nd_deriv}
 \partial^V_{y, \nu} \partial^V_{x, \mu} \tilde{\mathcal{D}}_{ov}
 = 
 \frac{2}{2 - \mathcal{D}_{ov}}
 \partial^V_{y, \nu} \partial^V_{x, \mu} \mathcal{D}_{ov}
 \frac{2}{2 - \mathcal{D}_{ov}}
 + \nonumber \\ +
 \frac{2}{2 - \mathcal{D}_{ov}} \partial^V_{x, \mu}\mathcal{D}_{ov}
 \frac{1}{2 - \mathcal{D}_{ov}} \partial^V_{y, \nu}\mathcal{D}_{ov}
 \frac{2}{2 - \mathcal{D}_{ov}}
 + \nonumber \\ +
 \frac{2}{2 - \mathcal{D}_{ov}} \partial^V_{y, \nu}\mathcal{D}_{ov}
 \frac{1}{2 - \mathcal{D}_{ov}} \partial^V_{x, \mu}\mathcal{D}_{ov}
 \frac{2}{2 - \mathcal{D}_{ov}}
\end{eqnarray}

 The first derivative $\partial^V_{x, \mu} \, \mathcal{D}_{ov}\lr{\mu_V}$ which enters (\ref{gw_proj_1st_deriv}) has been already calculated and is given by (\ref{overlap_derivative_vector2}). Differentiating this result once again over $V_{y, \nu}$ and using again the relations (\ref{revec_deriv_full}), (\ref{levec_deriv_full}) and (\ref{eval_deriv}), we obtain an expression for $\partial^V_{y, \nu} \partial^V_{x, \mu} \, \mathcal{D}_{ov}\lr{\mu_V}$ which is very similar to (\ref{overlap_derivative_vector_axial}) except for the factor $\lr{1 - s_i s_j}$ which is now replaced by $\lr{s_i - s_j}$.

 In order to complete the expressions (\ref{overlap_derivative_vector2}), (\ref{overlap_derivative_axial}) and (\ref{overlap_derivative_vector_axial}), we also give here explicit formulas for the operator $\mathcal{H} = \gamma_5 \mathcal{D}_w$ at finite chemical potential $\mu_V$ as well as for its derivatives over the vector gauge field and its eigenspectrum. In the background of an Abelian vector lattice gauge field $V_{x, \mu}$ the operator $\mathcal{H}$ reads
\begin{eqnarray}
\label{WDOperator}
 \mathcal{H}_{xy} = \lr{4 + \rho} \delta_{xy} \gamma_5
 - \nonumber \\ -
 \sum\limits_{\mu}\frac{\gamma_5 \lr{1 - \gamma^{\mu}}}{2} \, e^{i V_{x, \mu} + \delta_{\mu, 0} \mu_V} \, \delta_{y, x + \hat{\mu}}
 - \nonumber \\ -
 \sum\limits_{\mu}\frac{\gamma_5 \lr{1 + \gamma^{\mu}}}{2} \, e^{-i V_{x-\hat{\mu}, \mu} - \delta_{\mu, 0} \mu_V} \, \delta_{y, x - \hat{\mu}}  ,
\end{eqnarray}
where $x$, $y$ are the indices of lattice sites, $\hat{\mu}$ denotes the unit lattice vector in the direction $\mu$ (so that $x\pm\hat{\mu}$ is the index of the lattice site separated from the site $x$ by one lattice spacing in the direction $\mu$) and $\rho$ is the Wilson-Dirac mass parameter. In this work we are only interested in the derivatives of $\mathcal{H}$ over the spatial components of the vector gauge field. It is convenient to express them in terms of the derivatives of a spinor field $\mathcal{H} \psi$, where $\psi$ does not depend on $V_{x, \mu}$:
\begin{eqnarray}
\label{WDDerivative1}
 \partial^V_{x, i} \lrs{\mathcal{H} \psi}\lr{z}
 = \nonumber \\ =
 i \frac{\gamma_5 \lr{1 + \gamma_{i}}}{2} \, \psi\lr{z - \hat{i}} \delta\lr{z - \hat{i}, x}
 - \nonumber \\ -
 i \frac{\gamma_5 \lr{1 - \gamma_{i}}}{2} \, \psi\lr{z + \hat{i}} \delta\lr{z , x}
\nonumber \\
\label{WDDerivative2}
 \partial^V_{x, i} \partial^V_{y, j} \lrs{\mathcal{H} \psi}\lr{z}
 = \nonumber \\ = \delta_{ij} \delta_{xy}
 \frac{1 + \gamma_{i}}{2} \, \psi\lr{z - \hat{i}} \delta\lr{z - \hat{i}, x}
 + \nonumber \\ +
 \delta_{ij} \delta_{xy} \frac{1 - \gamma_{i}}{2} \, \psi\lr{z + \hat{i}} \delta\lr{z , x}
\end{eqnarray}

 In the basis of plane waves $\psi_x\lr{k} = \frac{1}{\sqrt{V}} e^{i k_{\mu} x^{\mu}}$ the operator $\mathcal{H}$ is block-diagonal with respect to $k_{\mu}$. The elements on the diagonal can be written in the following chiral block form:
\begin{eqnarray}
\label{WDOperatorFourier}
 \mathcal{H}
 = \nonumber \\ =
 \left(
   \begin{array}{cc}
     \Delta                                & \dslash{k} + i \sin\lr{k_0 - i \mu_V} \\
     \dslash{k} - i \sin\lr{k_0 - i \mu_V} &                               -\Delta \\
   \end{array}
 \right)  ,
\end{eqnarray}
where
\begin{eqnarray}
\label{delta_def}
\Delta = 2 \sin^2\lr{\frac{k_0 - i \mu_V}{2}}  + 2 \sum\limits_{i = 1}^{3}\sin^2\lr{\frac{k_{\mu}}{2}} - \rho ,
\end{eqnarray}
$0 < \rho < 2$ is the negative Wilson mass term which is necessary to define the overlap operator and $\dslash{k} = \sum\limits_{i=1}^{3} \sigma_i \sin\lr{k_i}$.

 Eigenvalues of $\mathcal{H}$ are then given by the eigenvalues of the $4 \times 4$ matrix (\ref{WDOperatorFourier}) for all possible values of $k_{\mu}$:
\begin{eqnarray}
\label{g5dw_eigval}
 \lambda_{s} \equiv s \lambda = s \sqrt{\Delta^2 + \sin^2\lr{k_0 - i \mu_V} + k^2} ,
\end{eqnarray}
where $k^2 = \sum\limits_{i=1}^{3} \sin^2\lr{k_i}$ and $s = \pm 1$. For each value of $s$ the eigenvalue (\ref{g5dw_eigval}) is two times degenerate. The corresponding eigenvectors are:
\begin{eqnarray}
\label{g5dw_eigenvecs}
 \ket{R_{s, \sigma}} = \mathcal{N}_{s} \, \psi_x\lr{k} \,
 \left(
   \begin{array}{c}
     \frac{i \sin\lr{k_0 - i \mu_V} + \dslash{k}}{s \lambda - \Delta} \ket{\chi_{\sigma}} \\
      \ket{\chi_{\sigma}}
   \end{array}
 \right)
 \nonumber \\
 \bra{L_{s, \sigma}} = \mathcal{N}_{s} \, \bar{\psi}_x\lr{k} \,
 \left(
   \begin{array}{c}
     \bra{\chi_{\sigma}} \, \frac{-i \sin\lr{k_0 - i \mu_V} + \dslash{k}}{s \lambda - \Delta}\\
     \bra{\chi_{\sigma}}
   \end{array}
 \right)  ,
\end{eqnarray}
where $\mathcal{N}_{s} = \sqrt{\frac{1}{2} - \frac{\Delta}{2 s \lambda}}$ and $\bra{\chi_{\sigma}}$, $\sigma = \pm 1$ are the two orthonormal Weyl spinors with $\ket{\chi_{\sigma}} \bra{\chi_{\sigma'}} = \delta_{\sigma \sigma'}$.

 Successively using the expressions (\ref{vv_correlator_observable}), (\ref{av_correlator_observable}), (\ref{inverse_gw_proj_1st_deriv}), (\ref{inverse_gw_proj_2nd_deriv}), (\ref{gw_proj_1st_deriv}), (\ref{gw_proj_2nd_deriv}), (\ref{overlap_derivative_vector2}), (\ref{overlap_derivative_axial}) and (\ref{overlap_derivative_vector_axial}), one can explicitly calculate the vector-vector and the axial-vector correlators (\ref{vv_correlator_def}) and (\ref{av_correlator_def}) once the spectrum of the operator $\mathcal{H}$ is known. In the free case, the eigensystem of $\mathcal{H}$ is given by (\ref{g5dw_eigval}) and (\ref{g5dw_eigenvecs}). The derivatives of $\mathcal{H}$ over the vector gauge field are given by (\ref{WDDerivative1}) and (\ref{WDDerivative2}).

\section{Perturbative non-renormalization of the transverse form-factors of the vector-vector-axial correlator}
\label{apdx:trans_formfact_nonrenorm}

 In this Appendix, we prove that the expression $w_T^{\lr{+}}\lr{k^2, k^2, 0} = -\frac{1}{k^2}$ (equation (\ref{w_plus_cme}) in the main text) for the transverse form-factor $w_T^{\lr{+}}\lr{p^2, q^2, \lr{p + q}^2}$ of the vector-vector-axial correlator (\ref{vva_correlator_def}) in the limit $p \rightarrow -q$ is valid in perturbation theory. Our proof is based on the work \cite{Knecht:04:1}, where the following relations among the form-factors which enter the decomposition (\ref{VVA_general_decomposition}) were proven to hold perturbatively in massless QCD:
\begin{widetext}
\begin{eqnarray}
\label{eqs_knecht_general}
   w_T^{\lr{+}}\lr{         p^2, q^2, \lr{p + q}^2}
 + w_T^{\lr{-}}\lr{         p^2, q^2, \lr{p + q}^2}
 - w_T^{\lr{+}}\lr{\lr{p + q}^2, q^2,          p^2}
 - w_T^{\lr{-}}\lr{\lr{p + q}^2, q^2,          p^2} = 0
 \nonumber \\
   \tilde{w}_T\lr{         p^2, q^2, \lr{p + q}^2}
 + w_T^{\lr{-}}\lr{         p^2, q^2, \lr{p + q}^2}
 + \tilde{w}_T\lr{\lr{p + q}^2, q^2,          p^2}
 + w_T^{\lr{-}}\lr{\lr{p + q}^2, q^2,          p^2} = 0
 \nonumber \\
                                            w_T^{\lr{+}}\lr{           p^2, q^2, \lr{p + q}^2}
 +                                          \tilde{w}_T\lr{           p^2, q^2, \lr{p + q}^2}
 +                                          w_T^{\lr{+}}\lr{\lr{p + q_2}^2, q^2,            p^2}
 +                                          \tilde{w}_T\lr{\lr{p + q_2}^2, q^2,            p^2}
 \nonumber \\
 + \frac{2 \lr{p + q} \cdot q}{p^2} w_T^{\lr{+}}\lr{\lr{p + q}^2, q^2, p^2}
 - \frac{2 p \cdot q}{p^2}            w_T^{\lr{-}}\lr{\lr{p + q}^2, q^2, p^2}             =  w_L\lr{\lr{p+q}^2, q^2, p^2} .
\end{eqnarray}
We consider these equations for momenta $p$ and $q$ which have equal squares $p^2 = q^2 = k^2$ and denote $\lr{p + q}^2 = \epsilon^2$. From the anti-symmetry of the form-factors $w_T^{\lr{-}}\lr{ p^2, q^2, \lr{p+q}^2} = -w_T^{\lr{-}}\lr{ q^2, p^2, \lr{p+q}^2}$ and $\tilde{w}_T\lr{ p^2, q^2, \lr{p + q}^2} = - \tilde{w}_T\lr{ q^2, p^2, \lr{p + q}^2}$ \cite{Knecht:04:1} it follows that $w_T^{\lr{-}}\lr{       k^2, k^2, \epsilon^2} = 0$ and $\tilde{w}_T\lr{ k^2, k^2, \epsilon^2} = 0$. Here we assume that all the form-factors are nonsingular in the limit $p^2 = q^2$. Indeed, since in this limit all the momenta entering the vector-vector-axial correlator (\ref{vva_correlator_def}) are still nonzero and non-collinear, there is no physical reason to expect such a singularity. We also take into account the explicit expression $w_L\lr{p^2, q^2, \lr{p+q}^2} = -\frac{2}{\lr{p+q}^2}$ for the longitudinal form-factor, which is completely fixed by the anomalous axial Ward identities \cite{Knecht:04:1}. Now the equations (\ref{eqs_knecht_general}) simplify to
\begin{eqnarray}
\label{eqs_knecht_kke}
   w_T^{\lr{+}}\lr{       k^2, k^2, \epsilon^2}
 - w_T^{\lr{+}}\lr{\epsilon^2, k^2,        k^2} = 0  ,
 \nonumber \\
    \tilde{w}_T\lr{\epsilon^2, k^2,        k^2}
 + w_T^{\lr{-}}\lr{\epsilon^2, k^2,        k^2} = 0  ,
 \nonumber \\
    w_T^{\lr{+}}\lr{        k^2, k^2, \epsilon^2}
 +  w_T^{\lr{+}}\lr{ \epsilon^2, k^2,        k^2}
 +   \tilde{w}_T\lr{ \epsilon^2, k^2,        k^2}
 \nonumber \\
 + \frac{\epsilon^2}{k^2} w_T^{\lr{+}}\lr{\epsilon^2, k^2, k^2}
 + \frac{2 k^2 - \epsilon^2}{k^2}            w_T^{\lr{-}}\lr{\epsilon^2, k^2, k^2}             =  - \frac{2}{k^2} .
\end{eqnarray}
Another set of equations can be obtained if we set $p^2 = \epsilon^2$, $q^2 = k^2$ and $\lr{p+q}^2 = k^2$ in (\ref{eqs_knecht_general}). Again taking into account that $w_T^{\lr{-}}\lr{       k^2, k^2, \epsilon^2} = 0$ and $\tilde{w}_T\lr{ k^2, k^2, \epsilon^2} = 0$, we obtain
\begin{eqnarray}
\label{eqs_knecht_ekk}
   w_T^{\lr{+}}\lr{ \epsilon^2, k^2,        k^2}
 + w_T^{\lr{-}}\lr{ \epsilon^2, k^2,        k^2}
 - w_T^{\lr{+}}\lr{        k^2, k^2, \epsilon^2} = 0  ,
 \nonumber \\
    \tilde{w}_T\lr{ \epsilon^2, k^2,        k^2}
 + w_T^{\lr{-}}\lr{ \epsilon^2, k^2,        k^2} = 0  ,
 \nonumber \\
   w_T^{\lr{+}}\lr{ \epsilon^2, k^2,        k^2}
 +  \tilde{w}_T\lr{ \epsilon^2, k^2,        k^2}
 + w_T^{\lr{+}}\lr{        k^2, k^2, \epsilon^2}
 \nonumber \\
 + \frac{2 k^2 - \epsilon^2}{\epsilon^2} w_T^{\lr{+}}\lr{k^2, k^2, \epsilon^2}  =  - \frac{2}{\epsilon^2}  .
\end{eqnarray}
\end{widetext}
We thus have six linear equations (\ref{eqs_knecht_kke}) and (\ref{eqs_knecht_ekk}) for the four unknowns $w_T^{\lr{+}}\lr{k^2, k^2, \epsilon^2}$, $w_T^{\lr{\pm}}\lr{\epsilon^2, k^2, k^2}$ and $\tilde{w}_T\lr{\epsilon^2, k^2, k^2}$. These equations are compatible and their unique solution is
\begin{eqnarray}
\label{eqs_knecht_solutions}
 w_T^{\lr{+}}\lr{k^2, k^2, \epsilon^2} = w_T^{\lr{+}}\lr{\epsilon^2, k^2, k^2} = - \frac{2}{2 k^2 + \epsilon^2}  ,
 \nonumber \\
 w_T^{\lr{-}}\lr{\epsilon^2, k^2, k^2} = \tilde{w}_T\lr{\epsilon^2, k^2, k^2} = 0  .
\end{eqnarray}
Taking now the limit $\epsilon^2 \rightarrow 0$, which corresponds to $p = k$, $q = -k$, we arrive at the expression $w_T^{\lr{+}}\lr{k^2, k^2, 0} = -\frac{1}{k^2}$.


\end{document}